\documentclass[modern]{aastex63}
\usepackage{microtype}  % ALWAYS!
\usepackage{amsmath}
\usepackage{amsfonts}
\usepackage{amssymb}
\usepackage{booktabs}
\usepackage{graphicx}
\usepackage{color}

\usepackage{enumitem}
\setlist[description]{style=unboxed}

\renewcommand{\twocolumngrid}{\onecolumngrid} % guess what this does HAHAHA!
\sloppypar

\newcommand{\nsources}{\ensuremath{232\,531}}
\newcommand{\Kmin}{M_{\rm min}}
\newcommand{\Kminval}{512}
\newcommand{\nbinary}{\ensuremath{19\,635}}

\newcommand{\goldsample}{\textit{Gold Sample}}
\newcommand{\ngold}{\ensuremath{1\,032}}
\newcommand{\nbimodal}{\ensuremath{492}}

\graphicspath{{}}
\newcommand{\documentname}{\textsl{Article}}
\newcommand{\sectionname}{Section}
\renewcommand{\figurename}{Figure}
\newcommand{\equationname}{Equation}
\renewcommand{\tablename}{Table}

\newcommand{\project}[1]{\textsl{#1}}

\newcommand{\acronym}[1]{{\small{#1}}}

\newcommand{\given}{\,|\,}
\newcommand{\norm}{\mathcal{N}}
\newcommand{\pdf}{\textsl{pdf}}

\newcommand{\dd}{\mathrm{d}}
\newcommand{\transpose}[1]{{#1}^{\mathsf{T}}}

\renewcommand{\vec}[1]{\ensuremath{\bs{#1}}}
\newcommand{\mat}[1]{\ensuremath{\mathbf{#1}}}

\newcommand{\msun}{\ensuremath{\mathrm{M}_\odot}}
\newcommand{\mjup}{\ensuremath{\mathrm{M}_{\mathrm{J}}}}
\newcommand{\kms}{\ensuremath{\mathrm{km}~\mathrm{s}^{-1}}}
\newcommand{\mps}{\ensuremath{\mathrm{m}~\mathrm{s}^{-1}}}

\newcommand{\dayd}{\ensuremath{\mathrm{d}}}
\newcommand{\yr}{\ensuremath{\mathrm{yr}}}
\newcommand{\AU}{\ensuremath{\mathrm{AU}}}

\newcommand{\bs}[1]{\boldsymbol{#1}}

\newcommand{\feh}{\ensuremath{{[{\rm Fe}/{\rm H}]}}}
\newcommand{\mh}{\ensuremath{{[{\rm M}/{\rm H}]}}}

\newcommand{\logg}{\ensuremath{\log g}}
\newcommand{\Teff}{\ensuremath{T_{\textrm{eff}}}}
\newcommand{\mtwomin}{\ensuremath{M_{2, {\rm min}}}}

\newcommand{\gaia}{\textsl{Gaia}}
\newcommand{\dr}[1]{\acronym{DR}#1}
\newcommand{\apogee}{\acronym{APOGEE}}
\newcommand{\sdss}{\acronym{SDSS}}
\newcommand{\sdssiv}{\acronym{SDSS-IV}}
\newcommand{\thejoker}{\project{The~Joker}}

\shorttitle{Close binaries in APOGEE DR16}
\shortauthors{Price-Whelan et al.}

\begin{document}

\title{Close Binary Companions to APOGEE DR16 Stars: \\
       20\,000 Binary-star Systems Across the Color--Magnitude Diagram}

\author[0000-0003-0872-7098]{Adrian~M.~Price-Whelan}
\affiliation{Center for Computational Astrophysics, Flatiron Institute,
             Simons Foundation, 162 Fifth Avenue, New York, NY 10010, USA}
\email{aprice-whelan@flatironinstitute.org}
\correspondingauthor{Adrian M. Price-Whelan}

\author[0000-0003-2866-9403]{David~W.~Hogg}
\affiliation{Center for Computational Astrophysics, Flatiron Institute,
             Simons Foundation, 162 Fifth Avenue, New York, NY 10010, USA}
\affiliation{Center for Cosmology and Particle Physics,
             Department of Physics,
             New York University, 726 Broadway,
             New York, NY 10003, USA}
\affiliation{Max-Planck-Institut f\"ur Astronomie,
             K\"onigstuhl 17, D-69117 Heidelberg, Germany}

\author[0000-0003-4996-9069]{Hans-Walter~Rix}
\affiliation{Max-Planck-Institut f\"ur Astronomie,
             K\"onigstuhl 17, D-69117 Heidelberg, Germany}

% APOGEE:
\author[0000-0002-1691-8217]{Rachael~L.~Beaton}
\altaffiliation{Hubble Fellow}
\altaffiliation{Carnegie-Princeton Fellow}
\affiliation{Department of Astrophysical Sciences, Princeton University,
             4 Ivy Lane, Princeton, NJ~08544}
\affiliation{The Observatories of the Carnegie Institution for Science,
             813 Santa Barbara St., Pasadena, CA~91101}

\author[0000-0002-7871-085X]{Hannah~Lewis}
\affiliation{Department of Astronomy, University of Virginia,
             Charlottesville, VA 22904-4325, USA}

\author[0000-0002-1793-3689]{David~L.~Nidever}
\affiliation{Department of Physics, Montana State University,
             P.O. Box 173840, Bozeman, MT 59717-3840}
\affiliation{NSF’s National Optical-Infrared Astronomy Research Laboratory,
             950 North Cherry Ave, Tucson, AZ 85719}

% APOGEE alphabetical:
\author{Andr\'es~Almeida}
\affiliation{Instituto de Investigaci\'on Multidisciplinario en Ciencia y
             Tecnolog\'ia, Universidad de La Serena, Benavente 980,
             La Serena, Chile}

\author[0000-0003-1086-1579]{Rodolfo~Barba}
\affiliation{Departamento de Astronom\'ia, Facultad de Ciencias,
             Universidad de La Serena, Cisternas 1200, La Serena, Chile}

\author{Timothy~C.~Beers}
\affiliation{Department of Physics and JINA Center for the Evolution of the
             Elements, University of Notre Dame, Notre Dame, IN 46556, USA}

\author{Joleen~K.~Carlberg}
\affiliation{Space Telescope Science Institute, 3700 San Martin Dr,
             Baltimore MD 21218}

\author{Nathan~De~Lee}
\affiliation{Department of Physics, Geology, and Engineering Technology,
             Northern Kentucky University, Highland Heights, KY 41099}
\affiliation{Department of Physics and Astronomy, Vanderbilt University,
             VU Station 1807, Nashville, TN 37235, USA}

\author{Jos\'e~G.~Fern\'andez-Trincado}
\affiliation{Instituto de Astronom\'ia y Ciencias Planetarias,
             Universidad de Atacama, Copayapu 485, Copiap\'o, Chile}

\author[0000-0002-0740-8346]{Peter~M.~Frinchaboy}
\affiliation{Department of Physics \& Astronomy, Texas Christian University,
             Fort Worth, TX, 76129, USA}

% \author{Domingo~An\'ibal Garc\'ia-Hern\'andez}
\author{D.~A.~Garc\'ia-Hern\'andez}
\affiliation{Instituto de Astrof\'isica de Canarias (IAC), E-38205 La Laguna,
             Tenerife, Spain}
\affiliation{Universidad de La Laguna (ULL), Departamento de Astrof\'isica,
             E-38206 La Laguna, Tenerife, Spain}

\author[0000-0002-8179-9445]{Paul~J.~Green}
\affil{Center for Astrophysics | Harvard \& Smithsonian, 60 Garden Street,
       Cambridge, MA 02138, USA}

\author{Sten~Hasselquist}
\altaffiliation{NSF Astronomy and Astrophysics Postdoctoral Fellow}
\affiliation{Department of Physics and Astronomy, University of Utah,
             115 S. 1400 E., Salt Lake City, UT 84112, USA}

\author{Pen\'elope~Longa-Pe{\~n}a}
\affiliation{Centro de Astronom{\'i}a (CITEVA), Universidad de Antofagasta,
             Avenida Angamos 601, Antofagasta 1270300, Chile}

\author{Steven~R.~Majewski}
\affiliation{Department of Astronomy, University of Virginia,
             Charlottesville, VA 22904-4325, USA}

\author{Christian~Nitschelm}
\affiliation{Centro de Astronom{\'i}a (CITEVA), Universidad de Antofagasta,
             Avenida Angamos 601, Antofagasta 1270300, Chile}

\author{Jennifer~Sobeck}
\affiliation{Department of Astronomy, University of Washington, Box 351580,
             Seattle, WA 98195, USA}

\author[0000-0002-3481-9052]{Keivan~G.~Stassun}
\affiliation{Department of Physics and Astronomy, Vanderbilt University,
             VU Station 1807, Nashville, TN 37235, USA}

\author{Guy~S.~Stringfellow}
\affiliation{Center for Astrophysics and Space Astronomy,
             Department of Astrophysical and Planetary Sciences,
             University of Colorado, 389 UCB,Boulder, CO 80309-0389, USA}

\author{Nicholas~W.~Troup}
\affiliation{Department of Physics, Salisbury University, Salisbury, MD 21801}

\begin{abstract}\noindent
% Context
Many problems in contemporary astrophysics---from understanding the formation of
black holes to untangling the chemical evolution of galaxies---rely on knowledge
about binary stars.
% Aims
This, in turn, depends on discovery and characterization of binary companions
for large numbers of different kinds of stars in different chemical and
dynamical environments.
Current stellar spectroscopic surveys observe hundreds of thousands to millions
of stars with (typically) few observational epochs, which allows binary
discovery but makes orbital characterization challenging.
% Methods
We use a custom Monte Carlo sampler (\thejoker) to perform discovery and
characterization of binary systems through radial-velocities, in the regime of
sparse, noisy, and poorly sampled multi-epoch data.
We use it to generate posterior samplings in Keplerian parameters for
\nsources\ sources released in \apogee\ Data Release 16.
% Results
Our final catalog contains \nbinary\ high-confidence close-binary ($P \lesssim
\textrm{few}~\textrm{years}$, $a \lesssim \textrm{few}~\AU$) systems that show
interesting relationships between binary occurrence rate and location in the
color--magnitude diagram.
We find notable faint companions at high masses (black-hole candidates), at low
masses (substellar candidates), and at very close separations (mass-transfer
candidates).
We also use the posterior samplings in a (toy) hierarchical inference to measure
the long-period binary-star eccentricity distribution.
We release the full set of posterior samplings for the entire parent sample of
\nsources\ stars.
This set of samplings involves no heuristic ``discovery'' threshold and
therefore can be used for myriad statistical purposes, including hierarchical
inferences about binary-star populations and sub-threshold searches.
\end{abstract}

% \keywords{}

\section*{~}\clearpage
\section{Introduction} \label{sec:intro}

Binary star systems provide key context and constraints for nearly all subfields
in astrophysics \citep[e.g.,][]{Breivik:BAAS, Rix:BAAS}.
For two concrete examples, a measurement of the occurrence rate of stellar-mass
black holes in the Milky Way would enable new constraints on binary black hole
formation channels to explain merger events observed by \acronym{LIGO}
\citep{LIGO:BH1, LIGO:catalog}, and interpretation of spectroscopic observations
of high-redshift galaxies and their stellar populations depends on understanding
the impacts of binary star evolution on stellar population parameters
\cite[e.g.,][]{Eldridge:2017}.
One common need for all applications is improved constraints on the population
properties (e.g., period, eccentricity, and mass ratio distributions, occurrence
rates) of stellar multiplets and their variations with stellar type, chemistry,
and dynamical environment, especially at the extrema of these stellar
characteristics.
This has only been comprehensively done for a sample of a few hundred stars in
solar neighborhood \citep{Raghavan:2010}, for specific stellar types
\citep[e.g.,][]{Moe:2017}, or with imprecise statistics using large samples of
stars \citep{Badenes:2018}.

While this problem spans a huge range in timescales (from hours to millenia),
current or near-future stellar surveys (e.g., \textit{Gaia}, \apogee,
\acronym{LAMOST}, \acronym{SDSS-V}; \citealt{Gaia-Collaboration:2016,
Gaia-Collaboration:2018, Majewski:2017, Zhao:2012, Kollmeier:2017}) have the
capacity to deliver samples of binary stars and stellar companions orders of
magnitude larger than are presently known, throughout all stages of stellar
evolution.
Even with existing datasets, we now have large enough sample sizes and the
robust tools needed to perform hierarchical inferences to constrain much more
detailed models of binary star formation and evolution \citep[e.g.,][]{Moe:2018,
El-Badry:2019a}.
However, the most precise measurements of the binary star population properties
will benefit from joint analysis of all stellar surveys, which cover a range of
stellar types, wavelengths, and measurement techniques.

As a step towards large-scale population inference, we focus here on multi-epoch
spectroscopic data from the \apogee\ surveys.
This survey sequence has predominantly targeted red-giant stars (although with
\dr{16} there are now many main-sequence stars, see below).
Because of operational re-visit decisions to reach signal-to-noise thresholds,
the surveys deliver some time-domain information, although they were not
designed with binary-star characterization as highest priority.
The fundamental data are $R\sim 20\,000$ $H$-band spectroscopy taken with the
primary purpose of mapping the Milky Way in elemental abundances and kinematics.
This survey is not the perfect target for binary-star identification, but it
arguably has one of the best combinations of spatial reach around the Milky Way,
coverage of the color-magnitude diagram, and multi-epoch data.

The challenge of working with data that were \emph{not} taken primarily for
binary characterization is that the data are sparse, time baselines are
variable, and most individual systems are not characterized uniquely (in orbital
parameters).
Indeed, period-fitting tasks generically produce multi-modal likelihood
functions and posterior pdfs for periods, amplitudes, and phases, and the Kepler
problem is no different.
These multi-modal likelihood functions are a nightmare for optimization or
sampling.
% many of our best people have given their lives in that forest of modes.
For this reason we created a custom sampler \thejoker\ \citep{thejoker} that
performs brute-force rejection sampling using a large, initial prior sampling.
Because this sampler does not produce Markov Chains, but rather just uses dense
prior samples, it does not get ``stuck'' in local optima; It samples the full
parameter space, with zero autocorrelation among samples.
To proceed, however, the sampler makes strong assumptions, that the source is a
single-lined spectroscopic binary (SB1) with only one companion (for example).
But it generates full samplings over orbital parameters for arbitrarily sampled
radial-velocity data.

Importantly, we use \thejoker\ to sample \emph{every} \apogee\ target as if it
were an SB1 (see \sectionname~\ref{sec:discussion} for some discussion of the
implications of this assumption).
Discovery of which stars really do have companions then becomes a
post-processing step on the confidence of the characterization: This project
deliberately conflates discovery with characterization.
This has the added advantage that we deliver full posterior samplings even for
stars that aren't obviously in binary systems; these can be used for statistical
studies and sub-threshold searches when new data arrive (i.e., future surveys or
follow-up).
In what follows, we run \thejoker\ on all of \apogee\ \dr{16} (subject to some
quality cuts) and release the resulting catalog of posterior samples, along with
some summary metadata for the subsample with good, uniquely determined
companions and orbits.
We demonstrate the use of the samplings with a few examples of interesting
objects and a simple hierarchical probabilistic populations inference.

This \documentname\ is similar to \cite{Price-Whelan:2018} but should be viewed
as a strict replacement (rather than an improvement) of the catalogs and results
of that work: We have made substantial improvements to the methodology
(improvements to \thejoker, see Appendix~\ref{app:marginal-likelihood}), and the
\apogee\ data in \dr{16} are of higher quality, substantially larger in
volume, and have more observation epochs.

\section{Data} \label{sec:data}

We use spectroscopic data from data release 16 (\dr{16}) of the \apogee\ surveys
(\citealt{Majewski:2017, DR16}; J\"onsson et al., in prep.).
\apogee\ is a component of the Sloan Digital Sky Survey IV (\sdssiv;
\citealt{Gunn:2006, Blanton:2017}) and its main goal is to survey the chemical
and dynamical properties of stars across much of the Milky Way disk by obtaining
high-resolution ($R \sim 22,500$; \citealt{Wilson:2019}) infrared ($H$-band)
spectroscopy of hundreds of thousands of stars.
The primary survey targets are selected with simple color and magnitude cuts
\citep{Zasowski:2013, Zasowski:2017}, but the survey uses fiber-plugged plates
that cover only a small fraction of the by area, which leads to extremely
nonuniform coverage of the Galactic stellar distribution (see, e.g., Figure~1 in
\citealt{DR16}).

\dr{16} is the first \sdss\ data release to contain \apogee\ data observed with
a duplicate of the \apogee\ spectrograph on the 2.5m Ir\'en\'ee du Pont
telescope \citep{Bowen:1973} at Las Campanas Observatory, providing access to
targets in the southern hemisphere.
For the first time, this data release also contains calibrated stellar
parameters for dwarf stars (J\"onsson et al., in prep.).
These two facts mean that \dr{16} contains nearly 3 times more sources with
calibrated stellar parameters than the previous public data release, \dr{14}
(\citealt{Abolfathi:2017, Holtzman:2018}; see Section~4 of \citealt{DR16} for
many more details about \apogee\ \dr{16}).

Most \apogee\ stars are observed multiple times in separate ``visits'' that are
combined before the \apogee\ data reduction pipeline \citep{Nidever:2015,
Zamora:2015, ASPCAP} determines stellar parameters and chemical abundances for
each source.
While the visit spectra naturally provide time-domain velocity information about
sources (thus enabling searches for massive companions), studying stellar
multiplicity is not the primary goal of the survey:
The cadence and time baseline for a typical \apogee\ source is primarily
governed by trying to schedule a set number of visits determined by
signal-to-noise thresholds for the faintest targets in a given field.
A small number of fields (five) were designed specifically for companion studies
and have $>10$ visits spaced to enable binary system characterization.

While some past studies have made use of other fields with large numbers of
visits to study binary-star systems \citep{Troup:2016, Fernandez-Trincado:2019},
a consequence of this strategy is that the time resolution and number of visits
for the vast majority of \apogee\ sources in \dr{16} is not sufficient for fully
determining companion orbital properties, as illustrated below.
Still, the large number of targets in \apogee\ and the dynamic range in stellar
and chemical properties offers an exciting opportunity to study the
\emph{population} of binary star systems as a function of these intrinsic
properties, even if most individual systems are poorly constrained.
We have previously developed tools to enable such studies \citep{thejoker}, as
summarized in \sectionname~\ref{sec:methods} below.
Here, we describe quality cuts we apply to the \apogee\ \dr{16} catalogs before
proceeding, and modifications to the visit-level velocity uncertainties to
account for the fact that they are generally underestimated by the \apogee\ data
reduction pipeline.

\subsection{Quality cuts and defining a parent sample}

The primary goal of this \documentname\ is to produce a catalog of posterior
samplings in Keplerian orbital parameters for \emph{all} high-quality \apogee\
sources in \dr{16} with multiple, well-measured radial velocities.
We therefore impose a set of quality cuts to sub-select \apogee\ \dr{16} sources
by rejecting sources or visits using the following \apogee\
bitmasks (\citealt{Holtzman:2018}, J\"onsson et al., in prep.):
\begin{itemize}
    \item Source-level (\texttt{allStar}) \texttt{STARFLAG} must not contain
    \texttt{VERY\_BRIGHT\_NEIGHBOR}, \texttt{SUSPECT\_RV\_COMBINATION} (bitmask
    values: 3, 16)
    \item Source-level (\texttt{allStar}) \texttt{ASPCAPFLAG} must not contain
    \texttt{TEFF\_BAD}, \texttt{LOGG\_BAD}, \texttt{VMICRO\_BAD},
    \texttt{ROTATION\_BAD}, \texttt{VSINI\_BAD} (bitmask value: 16, 17, 18, 26,
    30)
    \item Visit-level (\texttt{allVisit}) \texttt{STARFLAG} must not contain
    \texttt{VERY\_BRIGHT\_NEIGHBOR}, \texttt{SUSPECT\_RV\_COMBINATION},
    \texttt{LOW\_SNR}, \texttt{PERSIST\_HIGH}, \texttt{PERSIST\_JUMP\_POS},
    \texttt{PERSIST\_JUMP\_NEG} (bitmask value: 3, 9, 12, 13, 16)
\end{itemize}
These bitmasks are designed to remove the most obvious data reduction or
calibration failures that would directly impact the visit-level radial velocity
determinations.
However, we later impose a stricter set of quality masks when showing results in
\sectionname~\ref{sec:gold-sample}.
After applying the above masks, we additionally reject any source with $<3$
visits.
Our final parent sample contains \nsources\ unique sources, selected from the
$437,485$ unique sources in all of \apogee\ \dr{16}.
Of the $\approx$$200,000$ sources removed, the vast majority were dropped
because they had $<3$ visits ($\approx$$17\,000$ were removed by the quality
cuts).

% Notebook: Figure-DR16-statistics.ipynb
\begin{figure}[!t]
\begin{center}
\includegraphics[width=0.7\textwidth]{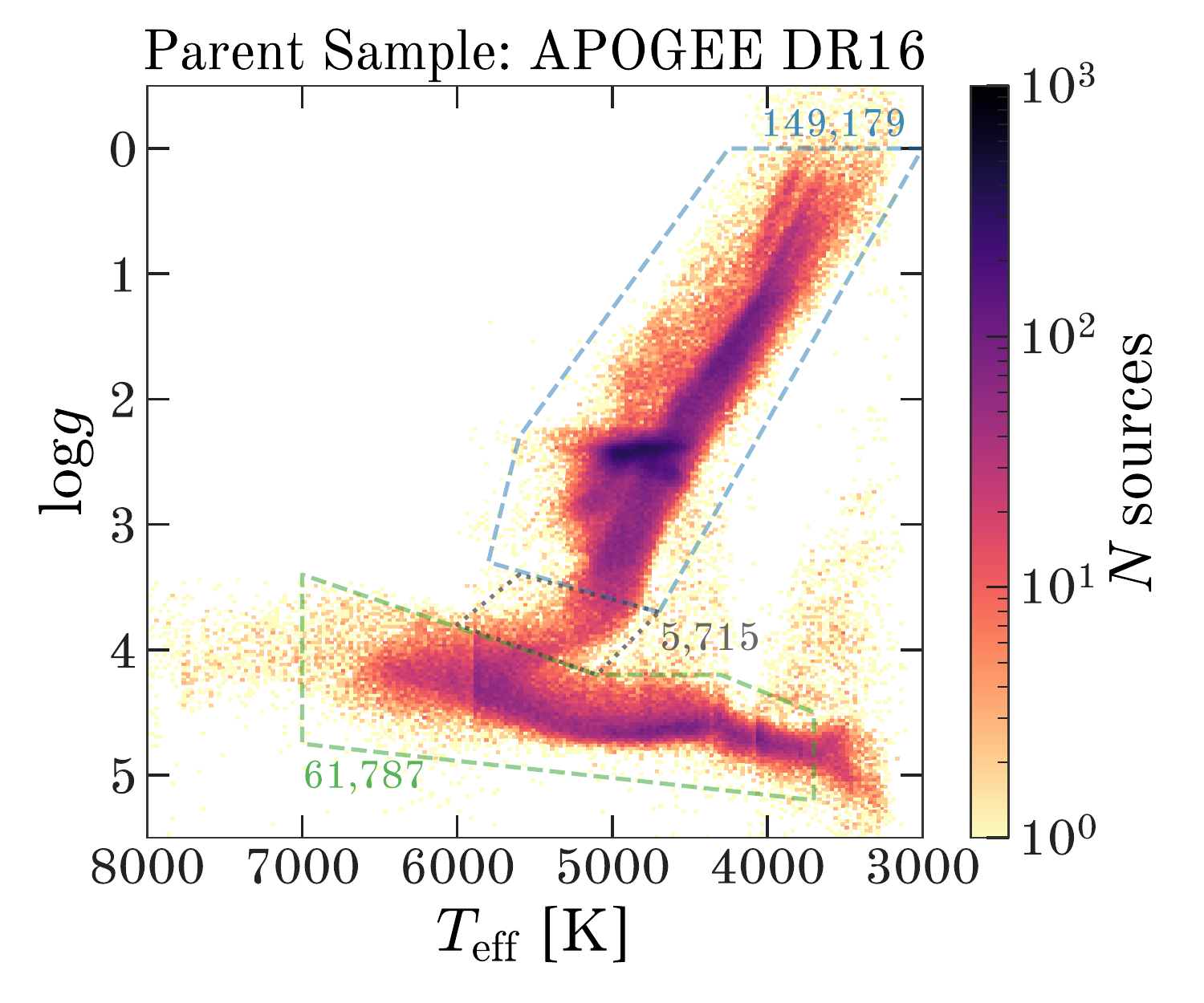}
\end{center}
\caption{%
Two spectroscopic (ASPCAP) stellar parameters---effective temperature, $T_{\rm
eff}$, and log-surface gravity, $\log g$---of \apogee\ \dr{16} sources that pass
our quality cuts.
These sources represent our ``parent sample.''
The pixel coloring indicates the number of sources in each bin of stellar
parameters.
The outlined regions roughly identify the red giant branch (upper polygon,
blue), subgiant branch (middle polygon, black), and (FGK-type) main sequence
(lower polygon, green).
The numbers next to each selection polygon indicate the number of sources in
each.
\label{fig:specHR}
}
\end{figure}

\begin{figure}[!t]
\begin{center}
\includegraphics[width=0.9\textwidth]{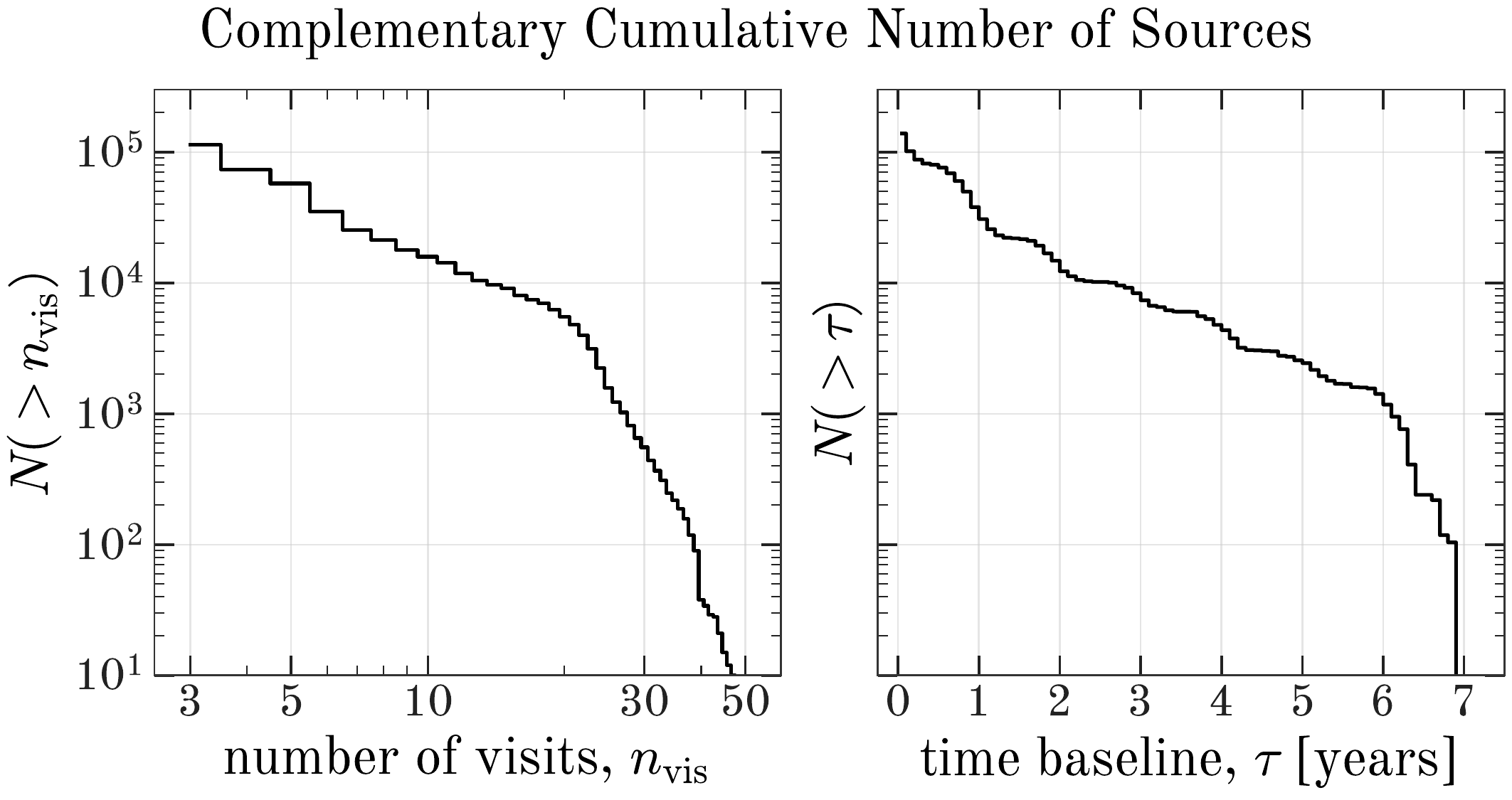}
\end{center}
\caption{%
Some statistics of \apogee\ \dr{16} visits.
\textbf{Left:} The number of sources with more than a given number of visits,
$n_{\rm vis}$.
While $\approx$$50\%$ of sources have 3 visits, ($114\,263$, $57\,593$,
$15\,862$) sources have $> (3, 5, 10)$ visits, respectively.
A very small number of sources have $>50$ visits.
\textbf{Right:} The number of sources with a time baseline, $\tau$, longer than
given (on the horizontal axis).
While $\approx$$50\%$ of sources have a time baseline $\tau \lesssim 56~{\rm
day}$, ($88\,737$, $9\,743$) sources have $\tau > (100, 1\,000)~{\rm days}$.
\label{fig:visitstats}
}
\end{figure}

\figurename~\ref{fig:specHR} shows the sources in our parent sample---i.e.
\apogee\ sources with 3 or more visits that pass the quality cuts described
above---as a function of spectroscopic stellar parameters $T_{\rm eff}$,
effective temperature, and $\log g$, log-surface gravity.
While the majority of sources are giant-branch stars ($>150\,000$), a
substantial number of main sequence stars are present ($>60\,000$) thanks to the
\apogee\ data reduction pipeline improvements for \dr{16} (J\"onsson et al., in
prep.).
Figure~\ref{fig:visitstats} shows some statistics about the time coverage of the
visits for sources in our parent sample.
About half of the sources have a small number of visits spread over a small time
baseline (the time spanned from the first to last visit for each source): $50\%$
of sources have $<5$ visits over $<100~{\rm days}$.
About $7\%$ of sources ($15\,366$) have $\geq 10$ visits over $\geq 100~{\rm
days}$.

% Notebook: Figure-DR16-statistics.ipynb

\subsection{Visit velocity uncertainty calibration} \label{sec:visitcalib}

The significance of apparent radial velocity variations, especially when
considering low-mass or long-period companions, will depend strongly on the
accuracy of the visit velocity measurement errors.
However, the catalog-level \apogee\ visit velocity errors (\texttt{VRELERR} in
the \texttt{allVisit} file) are known to be underestimated
\citep[e.g.,][]{Cottaar:2014}.
Here, we adopt the relation defined in Brown et al. (in prep.) to scale up the
visit velocity errors
\begin{equation}
    \sigma_v^2 = (3.5\,(\texttt{VRELERR})^{1.2})^2 + (0.072~\kms)^2 \label{eq:verr}
\end{equation}
where $\sigma_v$ is the adopted visit velocity error for a given visit, and
\texttt{VRELERR} is the uncertainty reported in the \apogee\ \dr{16} catalog.
This effectively applies a floor to the visit velocity errors of $72~\mps$ and
globally increases the error values by a multiplicative factor.

\section{Methods} \label{sec:methods}

As illustrated above, a large number of sources in \apogee\ \dr{16} have few
visits that span a short time baseline.
For most sources, we therefore expect that even if visit-level radial velocity
variations are detected with high significance, the companion orbital parameters
will be very uncertain---i.e. the posterior probability distribution function
(\pdf) over orbital parameters will generally be multi-modal with many
modes of comparable integrated probability \citep[e.g.,][]{thejoker}.
Still, in unison, or within the context of a hierarchical model that utilizes
the individual posterior samplings, the combination of all of these individually
weakly-constrained binary star orbits provides information about the population
of binary stars.
We have previously defined and implemented a custom Monte Carlo sampler for
precisely this problem: \thejoker\ \citep{thejoker}.
\thejoker\ is designed to deliver converged posterior samplings over Keplerian
orbital parameters given radial velocity observations, even when the
observations are sparse or very noisy.
Its prior application to \apogee\ \dr{14} \citep{Price-Whelan:2018} resulted in
a released catalog of over $5,000$ binary star systems;
this \documentname\ and companion catalog is a successor to this previous work.

As before, we use a parametrization of the two-body problem in which the radial
velocity, $v$, of an observed star in a binary system (referred to as the
\emph{primary}, even if it is less massive than its companion) can be expressed
as:
\begin{equation}
    v(t) = K \, \zeta(t \,;\, P, e, \omega, M_0, t_0) + v_0 \label{eq:kepler}
\end{equation}
where $K$ and $v_0$ are linear parameters (in time $t$).
The other parameters---period $P$, eccentricity $e$, argument of periastron
$\omega$, reference time $t_0$, and mean anomaly at reference time $M_0$---enter
through the nonlinear function
\begin{equation}
    \zeta(t \,;\, P, e, \omega, M_0, t_0) =
        \cos\left(f + \omega\right) + e\,\cos\omega \label{eq:zeta}
\end{equation}
where $f$ is the true anomaly, and we always set the reference time, $t_0$, to
the minimum observation time for a given set of radial velocity observations.

\subsection{Updates to \thejoker}
\label{sec:joker-update}

Since our initial paper defining \thejoker, we identified a conceptual error in
the assumptions made about the prior over the linear parameters ($K$, $v_0$) in
\citet{thejoker}.
We previously assumed that adopting a sufficiently broad, Gaussian prior over
the linear parameters meant that we could ignore an explicit definition of this
prior.
In particular, this allowed us to drop any terms related to the prior over these
parameters in the marginal likelihood expression \citep[\equationname~11
in][]{thejoker}.
This assumption is not correct, and can lead to unexpected behavior when applied
to data that are very noisy or have a small time baseline compared to the
samples of interest.
We have rewritten the expression for the marginal likelihood that underlies
\thejoker\ in Appendix~\ref{app:marginal-likelihood}, based on the notation in
Hogg et al. (in prep.).

Another issue with the assumption in the original implementation of \thejoker\
is that the prior over the velocity semi-amplitude, $K$, was identical at all
period and eccentricity values.
This implies vastly different prior beliefs about the companion mass as a
function of orbital period: For example, a zero-mean Gaussian prior on $K$ with
a standard-deviation of $30~\kms$ transforms to reasonable prior beliefs about
companion mass at periods around $1~\yr$, but gives substantial prior
probability to companion masses $>100~\msun$ at periods $>10^4~\dayd$.
We therefore, by default, adopt a new (also Gaussian) prior on the
semi-amplitude with a variance, $\sigma^2_K$, that scales with the
period and eccentricity such that
\begin{equation}
    \sigma^2_K = \sigma^2_{K, 0} \, \left(\frac{P}{P_0}\right)^{-2/3} \,
        \left(1 - e^2\right)^{-1} \label{eq:sigK}
\end{equation}
where $\sigma_{K, 0}$ and $P_0$ are additional hyperparameters that must be
specified.
This new prior on $K$ has the advantage that it, at fixed primary mass, has a
fixed form in companion mass that does not depend on period or eccentricity.

We have also made a number of improvements to the Python implementation of the
sampler.\footnote{\url{https://github.com/adrn/thejoker}}
For example, the prior distributions over all parameters are now specified as
\texttt{pymc3} \citep{Salvatier2016} distribution objects, meaning that the
priors over the nonlinear Keplerian parameters ($P$, $e$, $\omega$, and $M_0$;
see \citealt{thejoker}) are now fully customizable.
Using \texttt{pymc3} also enables compatibility with more efficient Markov Chain
Monte Carlo methods such as Hamiltonian Monte Carlo (HMC), which is useful for
seamlessly transitioning from generating posterior orbit samples with \thejoker\
to HMC when the system parameters are highly constrained (as discussed below).

\begin{table}[t]
    \centering
    \setlength{\tabcolsep}{0.75em} % for the horizontal padding
    \begin{tabular}{ r l p{6cm} }
    \toprule
    \textbf{parameter} & \textbf{prior} & \textbf{description} \\
    \toprule
    \multicolumn{3}{l}{\footnotesize \textsl{nonlinear parameters}}\\
    \hline
    $P$ & $\ln P \sim \mathcal{U}(2, 16384)~\textrm{day}$ & period \\
    $e$ & $e \sim \textrm{Beta}(0.867, 3.03)$ & eccentricity \\
    $t_0$ & fixed (minimum time) & reference time \\
    $M_0$ & $M_0 \sim \mathcal{U}(0, 2\pi)~\textrm{rad}$ & mean anomaly at reference time \\
    $\omega$ & $\omega \sim \mathcal{U}(0, 2\pi)~\textrm{rad}$ & argument of pericenter \\
    $s$ & $\ln s \sim \mathcal{N}(\mu_y, \sigma_y^2)$ & extra ``jitter'' added in quadrature to each visit velocity error \\
    \hline
    \multicolumn{3}{l}{\footnotesize \textsl{linear parameters}}\\
    \hline
    % $K$ & $K \sim \mathcal{N}(0, \sigma_{K}^2)~\kms$ & velocity semi-amplitude, where $\sigma_K$ is given by \equationname~\ref{eq:sigK} with $\sigma_{K,0} = 30~\kms$ and $P_0 = 365~\dayd$. \\
    $K$ & $K \sim \mathcal{N}(0, \sigma_{K}^2)~\kms$ & velocity semi-amplitude, where $\sigma_K$ is given by \equationname~\ref{eq:sigK}\\
    & $\sigma_{K,0} = 30~\kms$ & (see \equationname~\ref{eq:sigK})\\
    & $P_0 = 365~\dayd$ & (see \equationname~\ref{eq:sigK})\\
    $v_0$ & $v_0 \sim \mathcal{N}(0, \sigma_{v_0}^2)~\kms$ & system barycentric velocity \\
    & $\sigma_{v_0} = 100~\kms$ & \\
    \bottomrule
    \end{tabular}
    \caption{Summary and description of parameters and prior \pdf s. $\textrm{Beta}(a, b)$ is the
    beta distribution with shape parameters $(a, b)$, $\mathcal{U}(a, b)$ the
    uniform distribution over the domain $(a, b)$, and $\mathcal{N}(\mu,
    \sigma^2)$ is the normal distribution with mean $\mu$ and variance
    $\sigma^2$.}
    \label{tbl:prior}
\end{table}

\section{Running \thejoker\ on \apogee\ \dr{16}} \label{sec:rundr16}

For each of the \nsources\ sources in the parent catalog of sources selected
from \apogee\ \dr{16} (\sectionname~\ref{sec:data}), we run \thejoker\
\citep{thejoker} generate posterior samplings over the Keplerian orbital
parameters, including an additional per-source uncertainty or ``jitter''
parameter that is added in quadrature to the (adjusted) \apogee\ visit velocity
uncertainties (see above).
We start by generating a cache of $100,000,000$ prior samples for the nonlinear
parameters generated from the prior \pdf\ summarized in
\tablename~\ref{tbl:prior} (top rows).
For each source, we iteratively read random blocks of samples out of this cache,
evaluate the marginal likelihood, and rejection sample to produce posterior
samplings in the nonlinear parameters.
We repeat this iterative process until we reach the total number of samples in
the cache, or until we obtain a requested number of prior samples, $\Kmin$; this
is an arbitrary parameter that we set to $\Kmin = \Kminval$ for this work.
In practice, this is done by parallelizing the sampling (over sources) and takes
about 8 hours to run on 720 cores on our local compute cluster (at the Flatiron
Institute) for the entire sample.

After this procedure, the sources are in one of two stages of completion:
Sources either have $\Kmin = \Kminval$ posterior samples ($227\,999$ sources)
and are complete, or $\Kmin < \Kminval$ posterior samples ($4\,496$ sources) and
more samples are needed.
For sources that require more samples, these can be split again into two
classes: Sources with unimodal samplings in period, and sources with multi-modal
samplings (following the criteria described in \citealt{thejoker}).
For incomplete sources with multi-modal samplings ($2\,787$ sources), these
would need to be re-run with a much larger number of prior samples to reach
$\Kminval$ posterior samples, but here we mark these sources as incomplete.
For incomplete sources with unimodal samplings ($1\,709$ sources), we use the
samples returned from \thejoker\ to initialize Hamiltonian Monte Carlo (HMC) to
continue generating posterior samples.
We use \texttt{pymc3} with the \emph{No-U-Turn Sampler} (NUTS; \citealt{NUTS}),
using the dense mass matrix tuning prescription implemented in
\texttt{exoplanet} \citep{exoplanet:exoplanet} and run 4 chains in parallel,
each for 1000 tuning steps and 4000 steps subsequently.
We use these 4 chains and \texttt{pymc3} to compute the Gelman-Rubin convergence
statistic, $\hat{R}$ \citep{Gelman:1992}, for each parameter for each source; If
$\hat{R} < 1.1$ for all parameters, we down-sample to \Kminval\ samples and mark
that source as MCMC-completed.
If the HMC sampling fails (e.g., chains diverge substantially), we fall back to
the sampling from \thejoker---this generally only happens for data with serious
systematic issues (e.g., one outlier visit velocity with an unreasonable
velocity measurement).

At this point, we now have up to \Kminval\ posterior samples of the parameters
listed in \tablename~\ref{tbl:prior} for most of the \nsources\ \apogee\ sources
in the parent sample.
These full samplings will be released as a Value-Added Catalog (VAC) with the
\sdss\ \dr{16}+ ``mini'' data release planned for July 2020.

\section{A catalog of binary stars} \label{sec:catalog}

For some science cases and exploration, it is useful to define catalogs of
systems with likely companions from the posterior samplings generated here, as
we illustrate below.
This requires making decisions and imposing hard cuts on the samplings returned
by the above procedure.
However, most of the orbital parameter samplings for most sources are highly
uncertain and multi-modal, and it is therefore not possible to define simple
cuts on physical parameters (e.g., companion mass) to produce a simple catalog.
In the previous iteration of this work \citep{Price-Whelan:2018}, we defined
cuts based on percentiles computed from the distribution of $\ln K$ values after
comparing to running the same pipeline on a control sample of data generated
with purely Gaussian noise properties and time information taken from the
\apogee\ \dr{14} visits.
Here, we use both a selection based on the posterior samples in $K$ and based on
a likelihood ratio comparing the Keplerian orbit model assumed by \thejoker\
with a (robust) constant-velocity model for each source.

Following \citet{Price-Whelan:2018}, we again compute the 1st-percentile values
of the posterior samples in $\ln K$ for each source and refer to these values as
$P_{1\%}(\ln K)$;
this amounts to an estimate of the 99\% confidence lower-limit on the (log)
velocity semi-amplitude.
We again also generate a simulated control sample of data to assess
contamination when making selections using this quantity.
For each \apogee\ source in our parent sample, we take the maximum \textsl{a
posteriori} sample returned from the procedure defined above and subtract the
orbit computed from this sample from the visit velocity data.
We then re-run \thejoker\ on all of the residual data, and compute $P_{1\%}(\ln
K)$ for each star in the control sample.
We find that $<5\%$ of the control sample pass a cut of $P_{1\%}(\ln K) > 0$
(i.e., a conservative cut to require that sources have a velocity semi-amplitude
$> 1~\kms$).

For each \apogee\ source, after generating the posterior samplings, we compute
and store the maximum (over posterior samples) \emph{unmarginalized}
log-likelihood value for the Keplerian orbit model, i.e., for one source with
$N$ visit velocity measurements $v_n$ at times $t_n$,
\begin{equation}
    \ln \hat{L}_1 = \sum_n^N \ln \mathcal{N}(v_n \given
        v(t_n \,;\, \bs{\theta}_{\mathrm{max}}), \sigma_n^2)
\end{equation}
where $\sigma_n$ are the adjusted visit velocity errors
(\equationname~\ref{eq:verr}) and $v(\cdot)$ is given by
\equationname~\ref{eq:kepler} and is evaluated using the parameters for the
maximum likelihood posterior sample, $\bs{\theta}_{\mathrm{max}}$.

% Notebook: Figure-binary-examples.ipynb
\begin{figure}[!t]
    \begin{center}
    \includegraphics[width=0.8\textwidth]{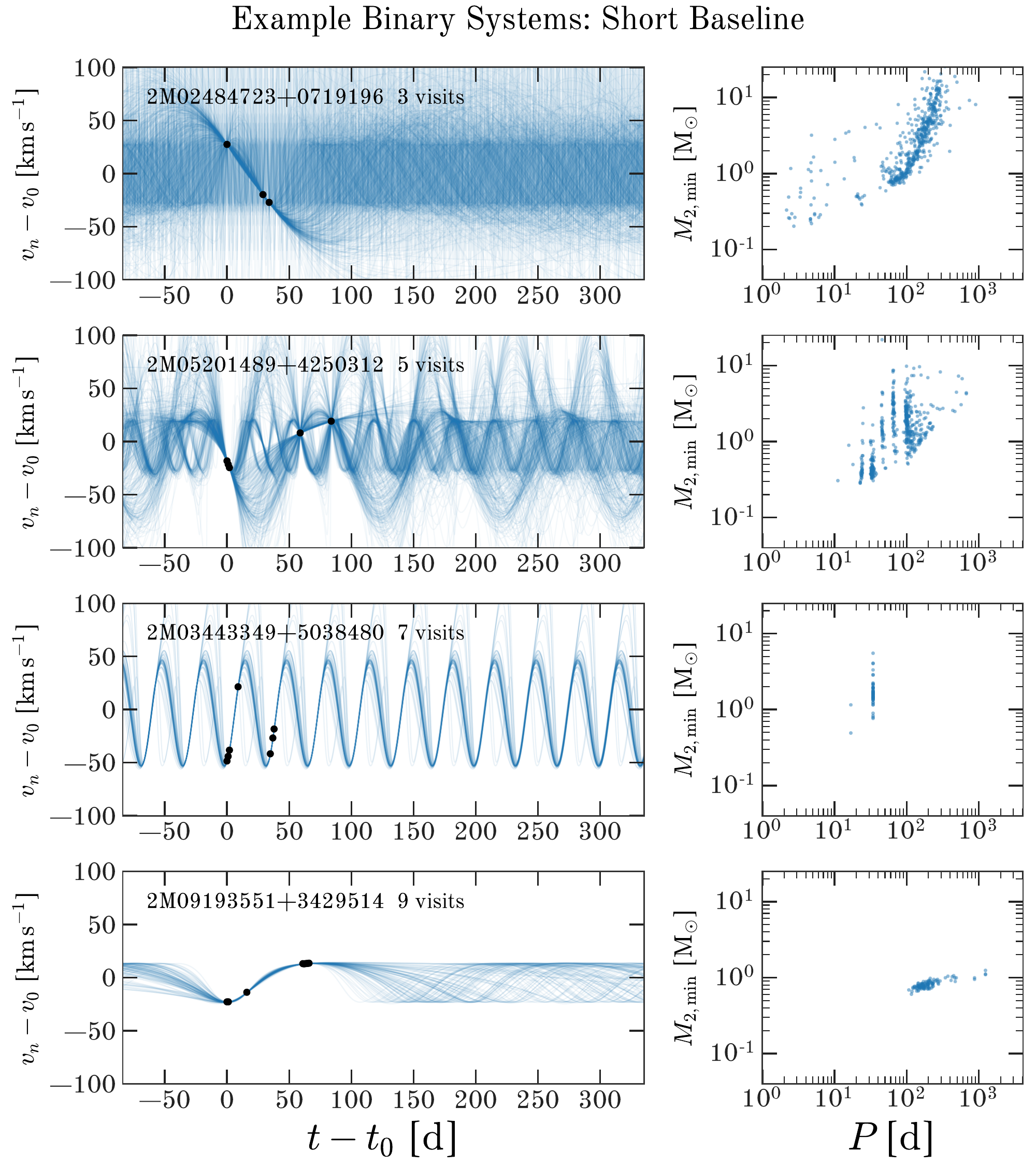}
    \end{center}
    \caption{%
    Example binary star systems that pass the selection and are included in the
    catalog released here for \apogee\ sources with short visit baselines
    ($\tau < 100~\dayd$).
    Each row is an \apogee\ source (indicated on the left panel).
    Left panels show the visit velocity data (black markers; error bars are
    typically smaller than the marker) and radial velocity orbits computed from
    the posterior samples (blue lines).
    Right panels show the same samples in period $P$ and minimum companion mass
    \mtwomin.
    \label{fig:binary-examples-short}
    }
\end{figure}

% Notebook: Figure-binary-examples.ipynb
\begin{figure}[!t]
    \begin{center}
    \includegraphics[width=0.8\textwidth]{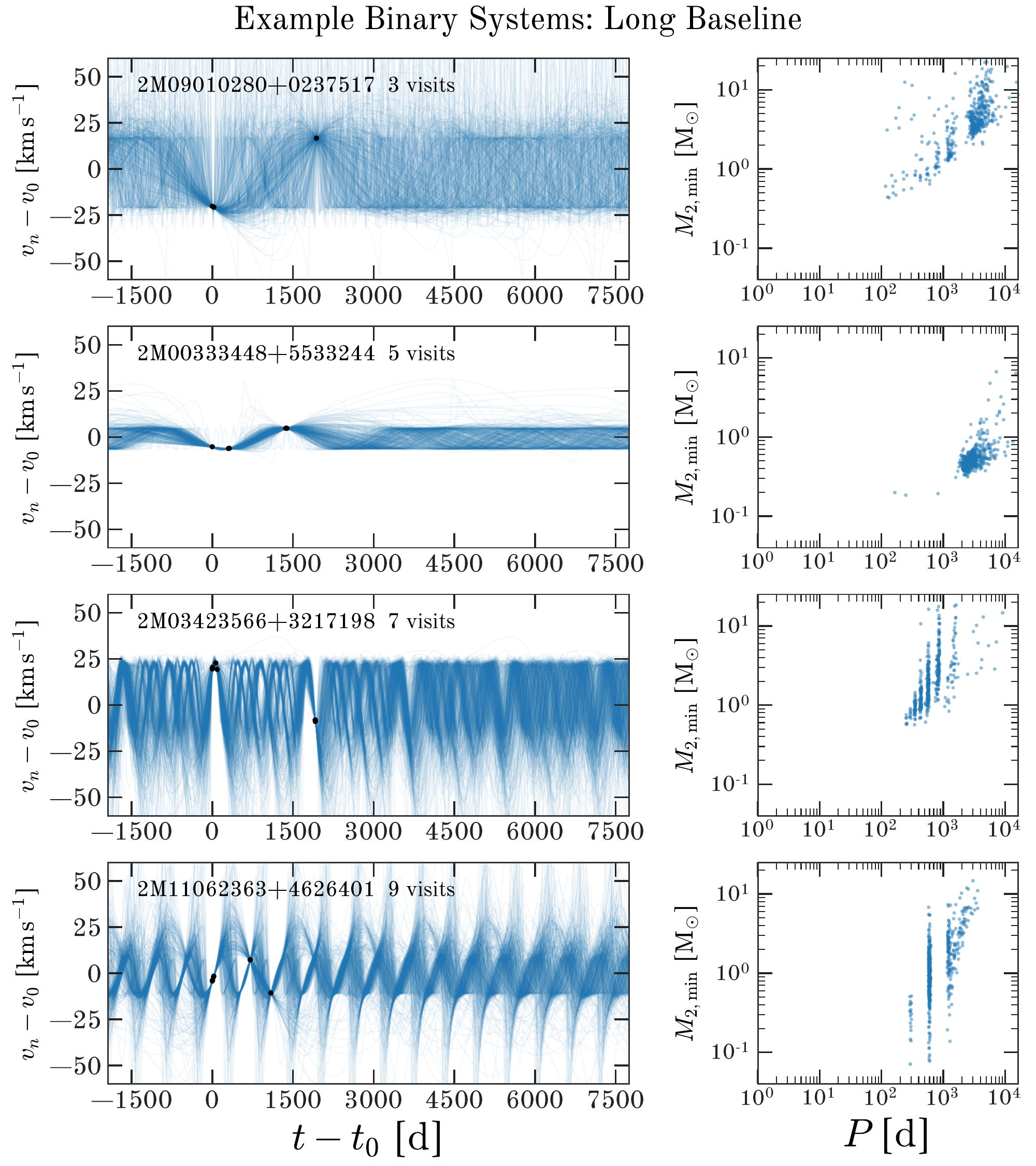}
    \end{center}
    \caption{%
    The same as \figurename~\ref{fig:binary-examples-short}, but for \apogee\
    sources with long visit baselines ($\tau > 1000~\dayd$)
    \label{fig:binary-examples-long}
    }
\end{figure}

For each source, we then also compute the maximum log-likelihood value for the
visit data under a model that assumes that the visit velocities are drawn from a
constant velocity with Gaussian uncertainties but allowing for $<20\%$ outliers;
we refer to this model as a \emph{robust constant-velocity} model for the visit
velocities.
Using the same notation for the visit velocity data as above, the likelihood
under this model is given by
\begin{equation}
    \hat{L}_2 = \prod_n^N \left[(1-f) \,
        \mathcal{N}(v_n \given \mu_v, \sigma_n^2)
        + f\, \mathcal{N}(v_n \given \mu_v, \sigma_n^2 + \Sigma) \right]
\end{equation}
where $f$ is the outlier fraction, $\mu_v$ is the constant-velocity value, and
$\Sigma$ is the variance of the outlier model component (assumed to be large).
We optimize this likelihood using the BFGS algorithm \citep{Nocedal:2006} with
bounds on the parameters such that $f \in (0, 0.2)$, $\mu_v \in (-500,
500)~\kms$, and $\Sigma \in (0, 3000)~(\kms)^2$.
We store the optimized log-likehihood value of the robust constant-velocity
model and refer to this as $\ln \hat{L}_2$

We define a catalog of binary star systems based on the posterior samplings
generated from \thejoker\ by selecting sources for which
\begin{align}
    P_{1\%}(\ln K) &> 0\\
    (\ln \hat{L}_1 - \ln \hat{L}_2) &> 4.6 \quad . \label{eq:binary-cuts}
\end{align}
The cut on the log-likelihood ratio comes from the (adopted) condition that
the maximum Kepler model likelihood should be $>100$ times the maximum robust
constant-velocity likelihood value.
Of the \nsources\ total sources in our parent sample, \nbinary\ sources pass the
selection above; those are binary systems where we can provide meaningful orbit
parameter samplings.
Summary information for the samplings generated from all sources in the parent
sample is included in \tablename~\ref{tbl:metadata}; in this table, the boolean
mask \texttt{binary\_catalog} can be used to select the \nbinary\ sources that
pass the binary star selections defined in \equationname~\ref{eq:binary-cuts}.

\figurename s~\ref{fig:binary-examples-short} and \ref{fig:binary-examples-long}
show some examples of binary systems that passed the selection above.
In each figure, each row is a different \apogee\ source, the two figures show
some example short baseline (\figurename~\ref{fig:binary-examples-short}) and
long baseline (\figurename~\ref{fig:binary-examples-long}) cases.
The left column of panels in each figure show the radial velocity data (black
markers) for randomly-chosen sources with (3, 5, 7, 9) visits (from top to
bottom), and the blue lines show radial velocity orbits computed from the
posterior samplings generated by \thejoker.
The right column of panels show the same posterior samples (blue markers), but
in a projection of the parameter space (period $P$ and minimum companion mass,
\mtwomin).
To compute \mtwomin, we use the posterior samplings from \thejoker\ along with
primary stellar masses computed by \citet{Queiroz:2019} by sampling over the
reported uncertainties on prior mass (assuming a Gaussian noise distribution on
primary mass).

\subsection{The Gold Sample}
\label{sec:gold-sample}

The majority of binary star systems that comprise the catalog defined above have
strongly multi-modal samplings in orbital properties.
While this is useful for binary \textit{population} studies, it is more
difficult to summarize the system orbital properties and their trends with
stellar properties as it is not possible to simply compress the samplings.
We therefore construct an additional catalog for the subset of sources that pass
a more stringent set of quality cuts and have converged, unimodal or bimodal
posterior samplings.
We define this \goldsample\ as sources that pass the following cuts:
\begin{itemize}
    \item Matches to a \gaia\ source within $2~\textrm{arcsec}$ of the
          reported \acronym{2MASS} sky position,
    \item Has a stellar mass measurement in the \acronym{STARHORSE} catalog
          \citep{Queiroz:2019},
    \item No additional \gaia\ sources within $2~\textrm{arcsec}$ with a
          $G$-band magnitude difference $\Delta G > -5$ (to remove sources that
          would lie within the \apogee\ fiber and appreciably contaminate the
          spectrum),
    \item No additional \gaia\ sources within $10~\textrm{arcsec}$ with a
          $G$-band magnitude difference $\Delta G > 2.5$ (to remove bright
          neighbor stars),
    \item $-0.5 < \logg < 5.5$ (reasonable stellar parameters),
    \item $3500 < \Teff < 8000$ (reasonable stellar parameters),
    \item $-2.5 < \mh < 0.5$ (reasonable stellar parameters),
    \item $s_{\textrm{MAP}} < 0.5~\kms$ (a small inferred excess variance),
    \item $n_{\textrm{vis}} > 5$ (more than 5 visit spectra).
\end{itemize}

We identify sources with unimodal posterior samplings as sources for which the
MCMC procedure succeeded (see \sectionname~\ref{sec:methods}).
We identify sources with bimodal samplings using the same procedure as
\cite{Price-Whelan:2018}.
Briefly, we use a $k$-means clustering algorithm with $k=2$ to identify clusters
of samples in orbital period, and assess whether the samplings in each cluster
are unimodal by checking whether all samples in each cluster lie within the mode
size defined in \cite{thejoker}.
In total, the \goldsample\ contains \ngold\ systems with unimodal samplings, and
\nbimodal\ systems with bimodal samplings.
Summary information and a list of sources in the \goldsample\ are included in
\tablename~\ref{tbl:goldsample}.

% Notebook: Figure-binary-fraction.ipynb
\begin{figure}[!t]
    \begin{center}
    \includegraphics[width=0.85\textwidth]{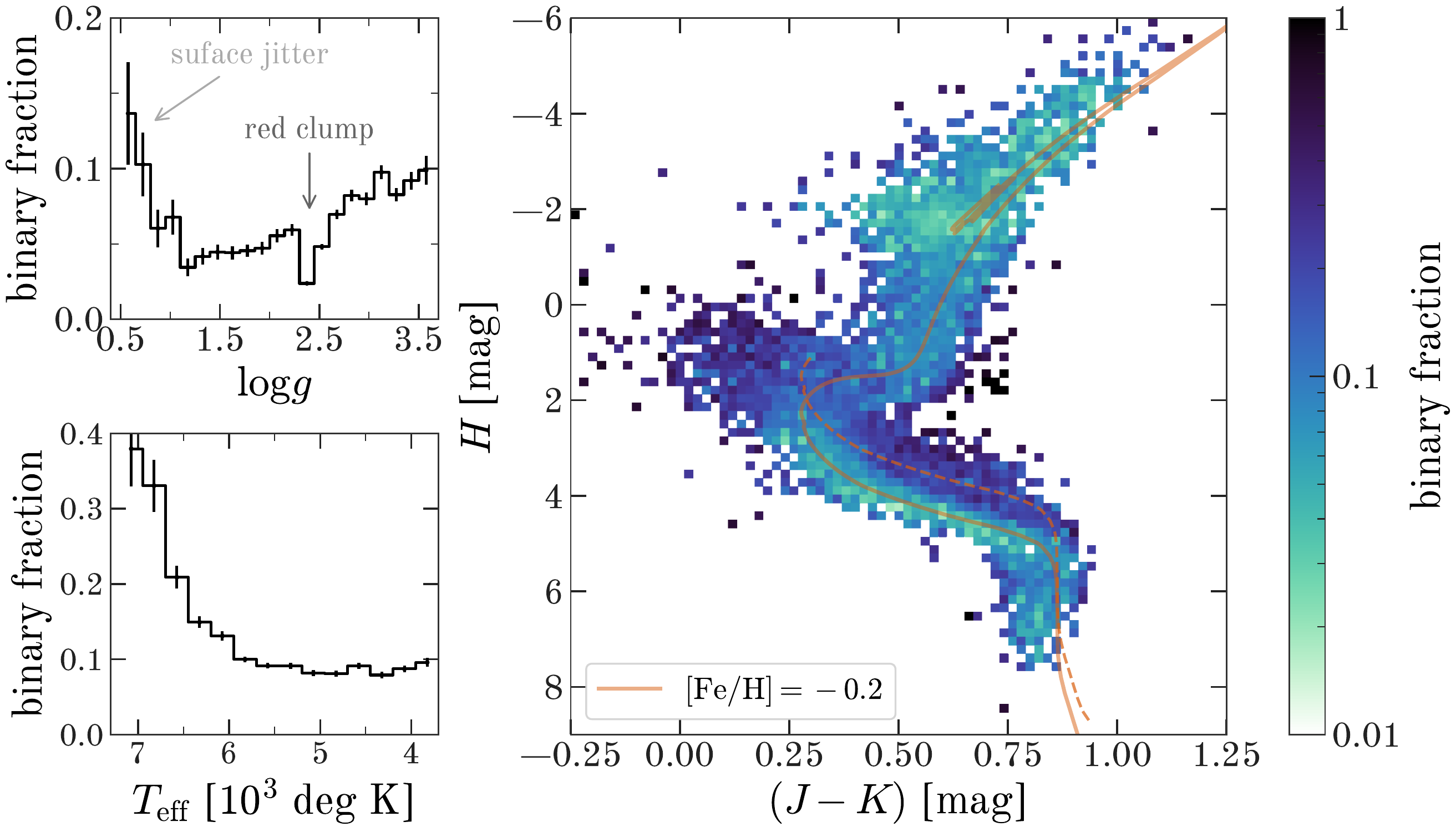}
    \end{center}
    \caption{%
    Close binarity across the color--magnitude diagram.
    \textbf{Upper left:} The observed close binary fraction as a function of
    spectroscopic surface gravity, $\log g$, for all stars that pass the
    giant branch selection indicated in \figurename~\ref{fig:specHR}.
    \textbf{Lower left:} The observed binary fraction as a function of
    spectroscopic effective temperature, $T_{\rm eff}$, for all stars that pass
    the main sequence selection indicated in \figurename~\ref{fig:specHR}.
    \textbf{Right:} The extinction-corrected \acronym{2MASS} color-magnitude
    diagram (CMD) for all \apogee\ sources, colored by the fraction of sources
    identified as binary-star systems (\sectionname~\ref{sec:catalog}).
    The solid (orange) line shows a \acronym{MIST} isochrone for a
    $5~\mathrm{Gyr}$ stellar population with $\feh = -0.2$, and the dashed line
    indicates the corresponding equal-mass binary sequence for main sequence
    stars.
    The panels in this figure are meant to be illustrative and should only be
    compared in a relative sense.
    \label{fig:binary-CMD}
    }
\end{figure}

\section{Results} \label{sec:results}

The epoch baselines for most \apogee\ sources, $\tau \lesssim 1~\yr$, imply that
most of the $\sim 20\,000$ binary systems (and certainly the $\sim 1\,000$ gold
sample systems) will have $P \lesssim \mathrm{years}$ and $a \lesssim
\mathrm{few}~\AU$.
However, the overall binary-star population extends from close binaries to
systems with $a\gtrsim 20\,000~\AU$, with a broad, approximately log-normal
period distribution centered at $\sim \log(250~\yr)$ \citep{Raghavan:2010}.
The binary systems we identified and study here are commonly referred to as
``close binaries'' \citep{Moe:2018, Badenes:2018}, representing the closest
$\sim 20\%$--$40\%$ of all bound binary star systems.
With this in mind, we use our sample to study some simply population properties
of these binary systems, and highlight some interesting systems in our sample.

\subsection{Close binary fraction trends with stellar properties}
\label{sec:binary-fraction}

Our catalog of binary stars is not complete in the sense that the cuts we have
made on the orbital parameter samplings will impart non-uniform selection biases
that depend on the true orbital properties of binaries and on the cadence of
observations (visits) for each source.
However, because of the simple target selection and observation strategy used by
\apogee\, \citep{Zasowski:2013, Zasowski:2017} and \apogee-2S (south; Beaton et
al., in prep., Santana et al., in prep.), we do not expect these binary-star
selection biases to depend strongly on stellar parameters (e.g., metallicity,
surface gravity).
We can therefore still use this catalog to study the \emph{relative} close
binary fraction within our sample, but caution against interpreting the binary
fractions discussed below in an absolute sense.
Note that here we use ``binary fraction'' to mean the observed fraction of
detected binary systems, not the intrinsic or birth fraction of binary systems.

\figurename~\ref{fig:binary-CMD} (right) shows a near-infrared, binned
color-magnitude diagram for all \apogee\ sources in the subset of our sample
that cross-match to the \gaia\ \dr{2} astrometric catalog
\citep{Gaia-Collaboration:2016, Gaia-Collaboration:2018, Gaia:2018a} and have a
parallax signal-to-noise $\varpi / \sigma_\varpi > 8$; we compute the absolute
$H$-band magnitudes by converting parallax into distance as $d = 1/\varpi$.
Each pixel is colored by the ratio of the number of binary star systems
identified by the selections defined above (\equationname~\ref{eq:binary-cuts})
over the total number of sources in the parent sample; Pixels with fewer than
four stars in the parent sample are white.
The solid (orange) line shows a $5~\mathrm{Gyr}$ \acronym{MIST} isochrone
\cite{Dotter:2016, Choi:2016, Paxton:2011, Paxton:2013, Paxton:2015} with
metallicity indicated in the figure legend.
The dashed line shows the equal-mass binary sequence for this isochrone,
$0.75~\mathrm{mag}$ above the main sequence.
Note that, as expected, the binary fraction is higher in this region of the CMD,
and is higher towards younger and more massive main sequence stars.
Note also that the red clump (around $(J-K) \approx 0.6$, $H \approx -1.8$) has
a \emph{low} binary fraction (as noted in \citealt{Badenes:2018}).
The width of the binary sequence (in $H$ magnitude) can be explained by the
spread in metallicities in the \apogee\ sample.

% Notebook: Figure-binary-fraction.ipynb
\begin{figure}[!t]
    \begin{center}
    \includegraphics[width=0.6\textwidth]{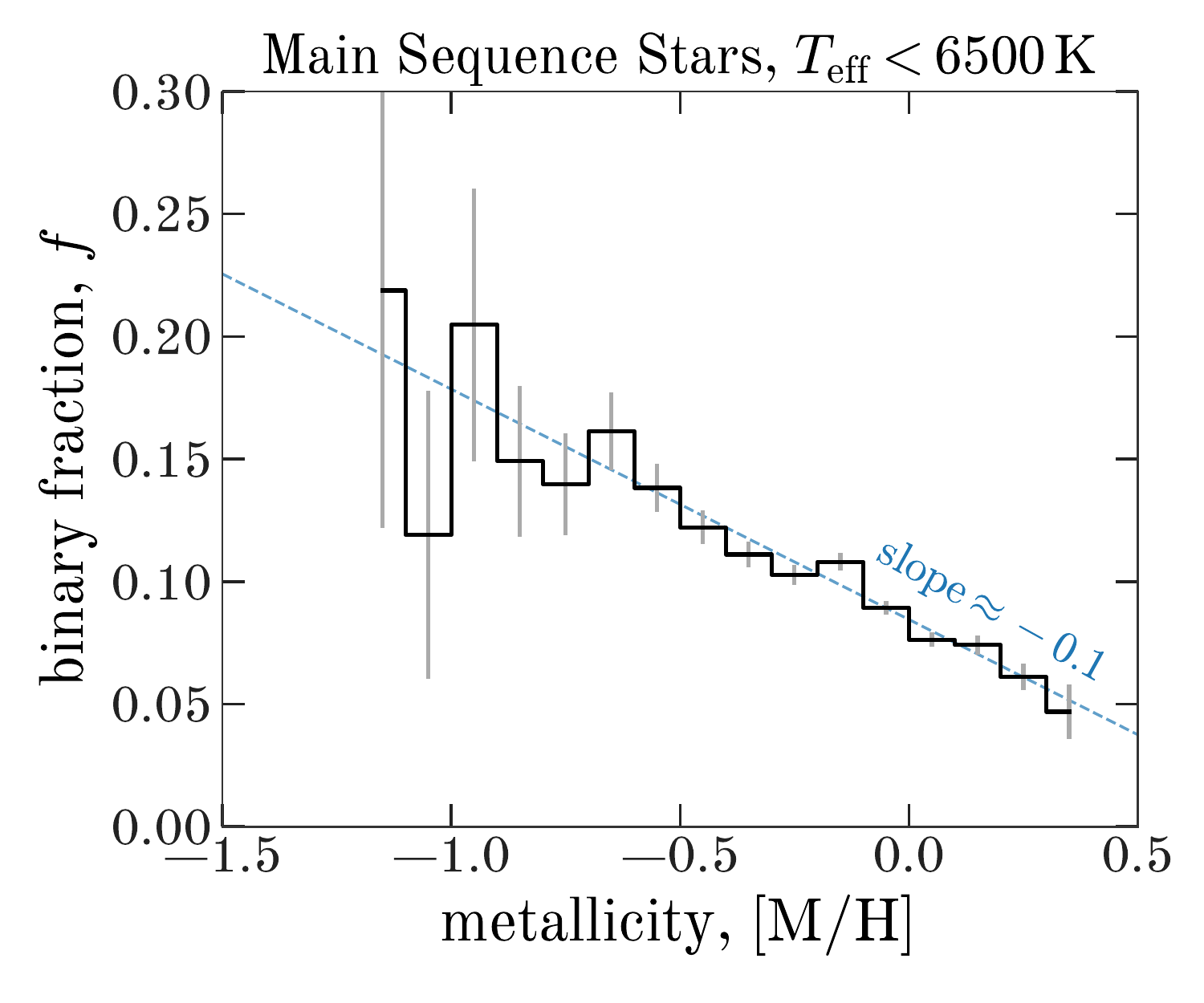}
    \end{center}
    \caption{%
    The observed close binary fraction as a function of bulk metallicity, $[{\rm
    M}/{\rm H}]$.
    The binary fraction is anti-correlated with metallicity, here measured with
    a slope of $-0.1$.
    \label{fig:binfrac-mh}
    }
\end{figure}

The top left panel of \figurename~\ref{fig:binary-CMD} shows another view of the
binary fraction, here shown as a 1D function of surface gravity for stars in the
giant branch selection region shown in \figurename~\ref{fig:specHR}.
There is a clear and sharp dip in the occurrence of binary systems near the red
clump, and the close binary fraction decreases appreciably with decreasing
surface gravity (i.e., increased size).
Both of these features are likely signatures of companion engulfment: As red
giant stars evolve up the giant branch, any companions with orbital semi-major
axes smaller than a few times the surface size of the primary star could be
consumed \citep{Ivanova:2013}, therefore leading to an overall decrease in the
binary fraction for larger stars.

The bottom left panel shows the same, but as a function of effective temperature
(as a proxy for the mass of the primary) for stars in the main sequence
selection region shown in \figurename~\ref{fig:specHR}.
As has been noted in many past studies, we find that the binary fraction
(companion frequency) increases with stellar mass (effective temperature) along
the main sequence \citep[e.g.,][]{Duchene:2013}.
Like the compilation of companion frequencies shown in \cite{Duchene:2013} and a
number of studies of local samples of binary systems
\citep[e.g.,][]{Eggleton:2008, Raghavan:2010, Gao:2014}, we find that the binary
fraction increases steeply for primary stellar masses $M_1 \gtrsim 1.1~\msun$
($\Teff \gtrsim 6000~\textrm{K}$).

\figurename~\ref{fig:binfrac-mh} shows the total close binary fraction for main
sequence stars as a function of bulk metallicity, $\mh$.
As has been recently emphasized, we find that the fraction of close binaries is
significantly anti-correlated with metallicity \citep{Moe:2019, El-Badry:2019a},
in disagreement with previous work (with a much smaller sample) that had
found no dependency with metallicity \citep{Jenkins:2015}.
We find a shallower dependence of these properties, with a slope of $\approx
-0.1$ as compared to the previously determined slope of $-0.2$ (also for close
binaries) \citep{Moe:2019}.
While we do not expect there to be strong selection biases that imprint on these
quantities, we emphasize that we have not corrected for detection efficiency or
completeness with our sample.

\subsection{Orbital parameter trends with stellar properties}
\label{sec:gold-sample}

% Notebook: Figure-gold-sample.ipynb
\begin{figure}[!t]
    \begin{center}
    \includegraphics[width=0.85\textwidth]{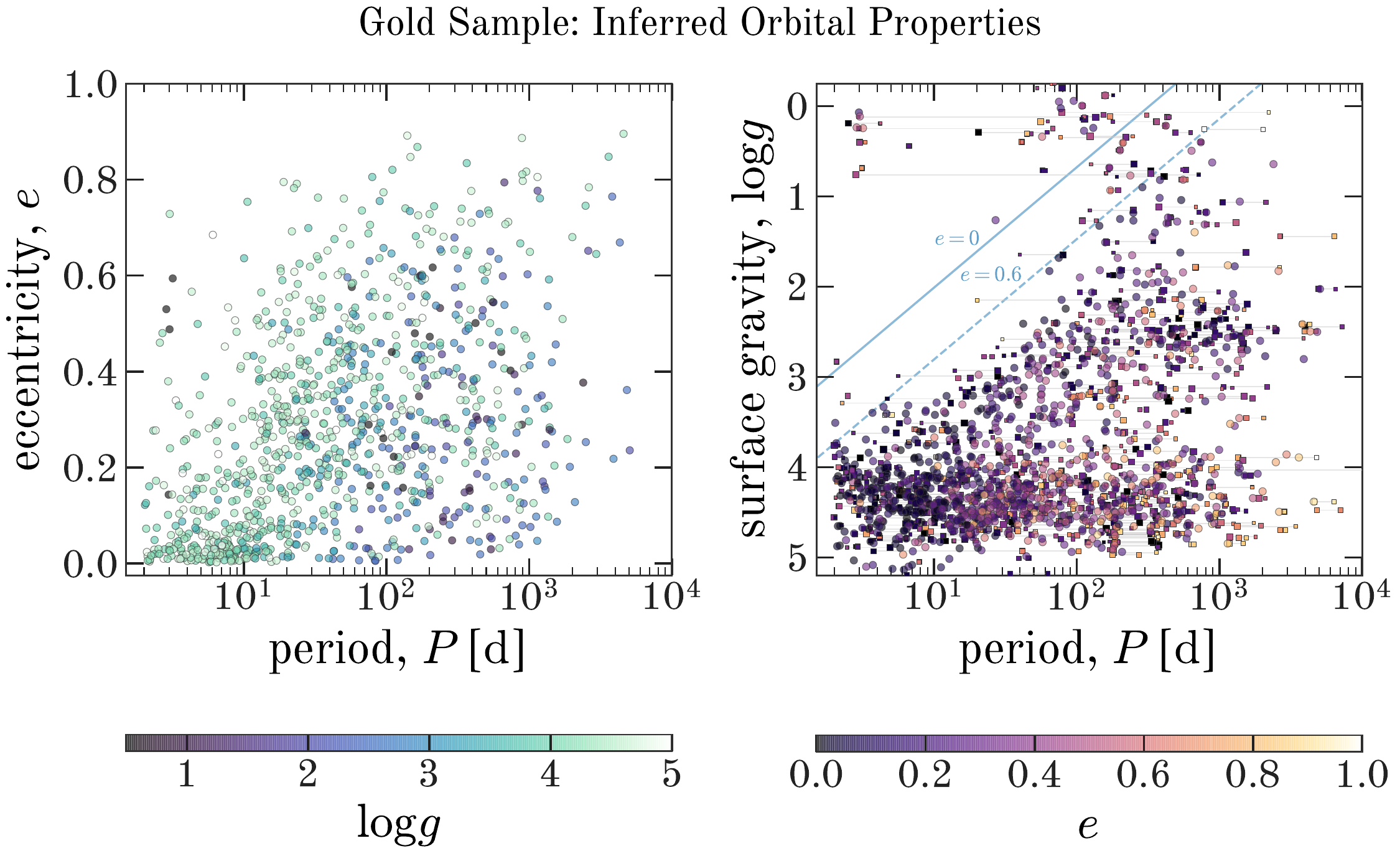}
    \end{center}
    \caption{%
    \textbf{Left:} Inferred orbital periods, $P$, and eccentricities, $e$, for
    \goldsample\ systems with unimodal samplings.
    Markers are colored by surface gravity, \logg.
    \textbf{Right:} Similar to the left panel, but showing orbital periods and
    \apogee\ \logg\ measurements.
    The circular markers again indicate the \ngold\ sources with unimodal
    samplings, and the square markers indicate the \nbimodal\ sources with
    bimodal samplings, where the mean of each mode is plotted and connected by a
    horizontal gray line and the relative sizes of the square markers indicate
    the fraction of the samples that lie in each mode.
    The diagonal (blue) lines show the orbital period of a $0.3~\msun$ companion
    with an orbital pericentric radius equal to the surface size of a
    $1.1~\msun$ star with the given surface gravity and labeled eccentricity,
    $e$.
    \label{fig:Plogg}
    }
\end{figure}

\figurename~\ref{fig:Plogg} shows inferred orbital periods for all stars in the
\goldsample\ as a function of orbital eccentricity (left) and surface gravity
(right), where the range of periods is clearly limited (at large $P$) by the
The left panel clearly shows the impact of tidal circularization
\citep[e.g.,][]{Zahn:1977, Meibom:2005}.
The eccentricities of systems with $P \lesssim 10~\dayd$ are much more peaked
near $e=0$ as compared to systems with larger orbital periods (e.g., $P\gtrsim
100~\dayd$).
However, for systems with giant-star members, circularization occurs at longer
periods \citep[e.g.,][]{Price-Whelan:2018a}: Systems with low eccentricities and
$P>30~\dayd$ tend to have lower \logg\ values (darker points) as compared to
systems with $P<10~\dayd$.
In this panel, we visually inspected the small number of systems with $P\lesssim
5~\dayd$ and $e \gtrsim 0.4$ and found that most of these cases appear to result
from systematic errors with the visit velocity data that were not removed by the
quality cuts.

The right panel of \figurename~\ref{fig:Plogg} shows orbital periods of systems
in the \goldsample\ (with unimodal or bimodal samplings) as a function of
surface gravity.
The diagonal lines in this panel show the orbital periods at which the orbital
pericenter of a $0.3~\msun$ companion is equal to the stellar surface radius for
a $1.1~\msun$ primary star (the median RGB mass in our sample) at the given
surface gravity and eccentricity indicated.
The apparent lack of systems with orbital periods shorter than or within
$\sim$1~$\mathrm{dex}$ these critical lines (with $\logg \gtrsim 1$) suggests
that a substantial fraction of binaries must merge or disrupt during the
evolution of the primary star.
The paradoxical points with seemingly small orbital periods at small \logg\
(i.e., suggesting they orbit within the surface of the primary star) are likely
due to contamination from asteroseismic modes that manifest as velocity
``jitter'' in low-surface-gravity giant stars.
This jitter can reach amplitudes of $>1~\kms$ already by $\logg \approx 1$ and
likely increases towards even lower values of \logg\
\citep[e.g.,][]{Hekker:2008}.

\begin{figure}[t]
    \begin{center}
    \includegraphics[width=0.95\textwidth]{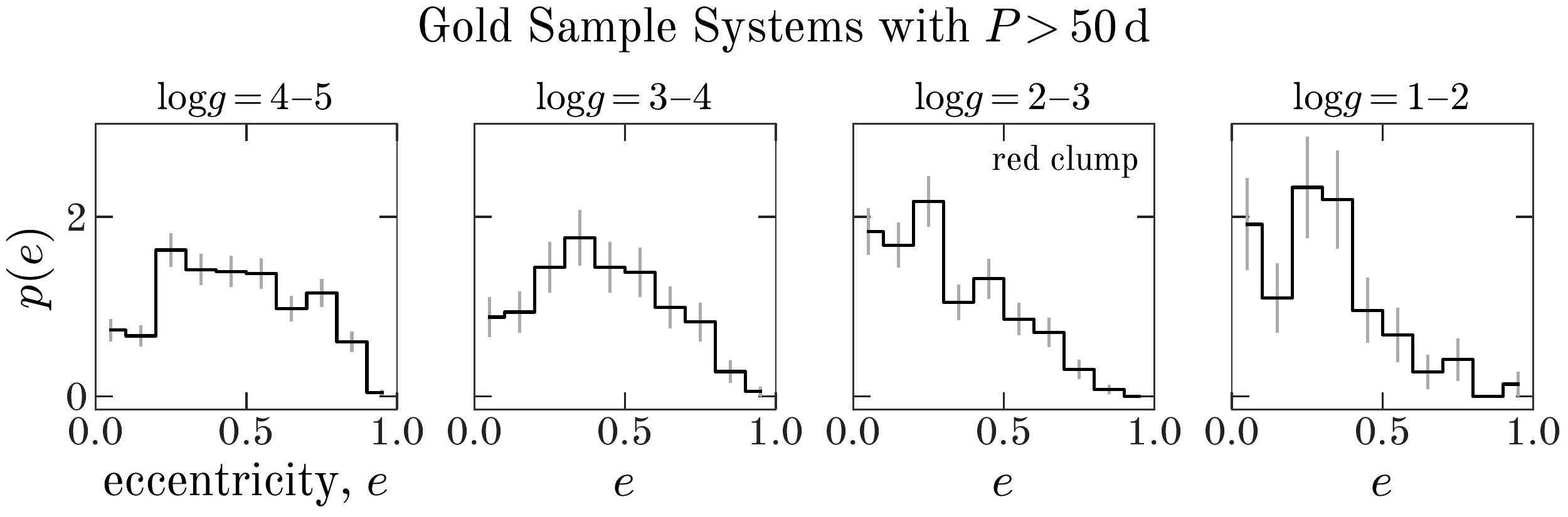}
    \end{center}
    \caption{%
    Histograms of maximum \textsl{a posteriori} (MAP) eccentricity values for
    systems with MAP $P > 50~\dayd$.
    Each panel contains systems with primary star surface gravities indicated in
    the title.
    \label{fig:Pe-logg-bins}
    }
\end{figure}

Also note the gradients in eccentricity (i.e., marker color) with respect to
orbital period in the right panel of \figurename~\ref{fig:Plogg}.
At a given surface gravity, there tend to be more black points (low
eccentricity) closer to the stellar surface (closer to the diagonal lines).
However, near the red clump ($2 \lesssim \logg \lesssim 3$), there appears to be
an over-abundance of low-eccentricity points at orbital periods $P\gtrsim
100~\dayd$.
\figurename~\ref{fig:Pe-logg-bins} shows histograms of maximum \textsl{a
posteriori} MAP eccentricity values for sources in the \goldsample\ with MAP
period values $P > 50~\dayd$ in four bins of surface gravity, from main sequence
(farthest left) to upper giant branch (farthest right).
Note that around the red clump (third panel from the left), there are
significantly more $e < 0.1$ systems than expected (i.e., as compared to the
previous bin).
We therefore posit that systems with a primary around the red clump that have
low eccentricities have likely already ascended the giant branch, whereas
systems with large eccentricities are likely on their first ascent up the giant
branch.

% Notebook: Figure-gold-sample.ipynb

\subsection{Interesting low and high mass companions}
\label{sec:low-high-mass}

\begin{figure}[t]
    \begin{center}
    \includegraphics[width=0.85\textwidth]{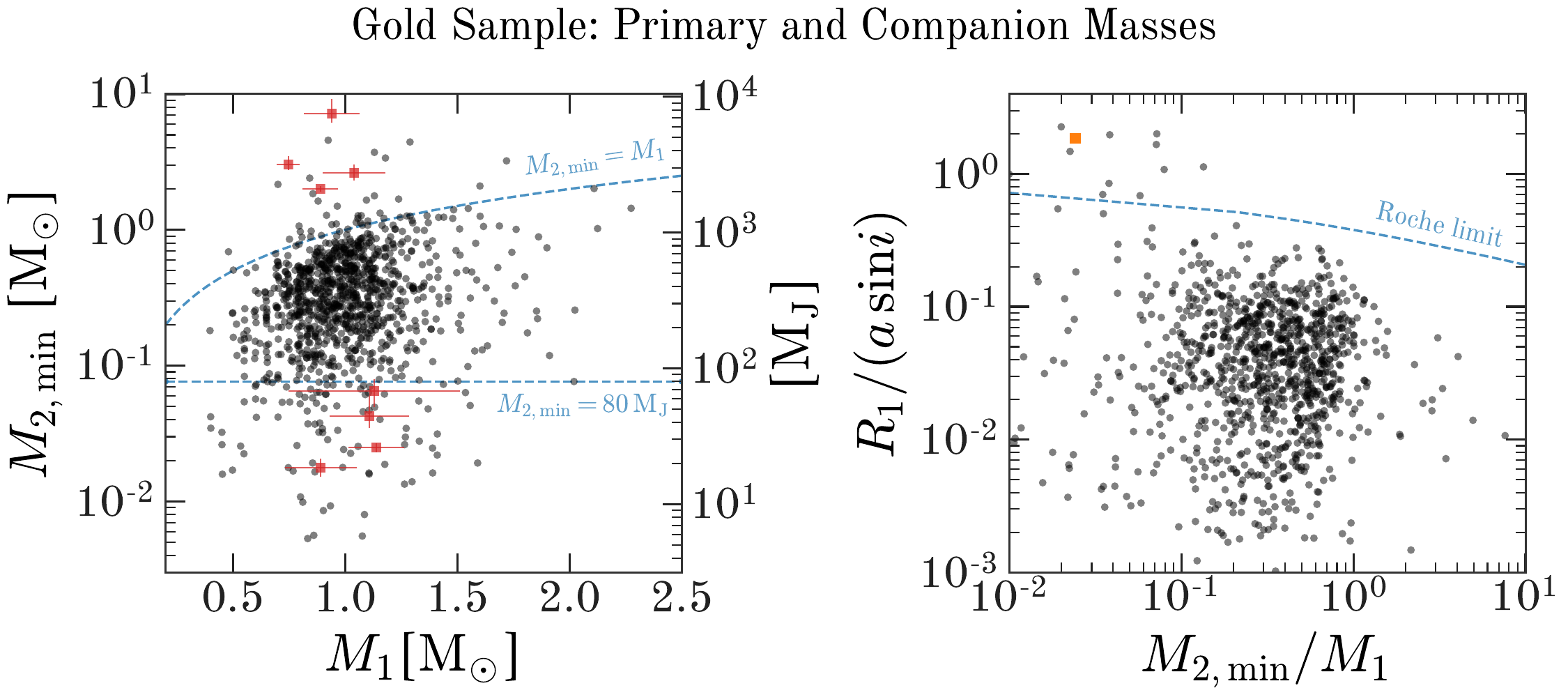}
    \end{center}
    \caption{%
    \textbf{Left:} Minimum companion masses, $\mtwomin$, as a function of
    primary mass, $M_1$, for all stars in the \goldsample.
    The upper dashed (blue) line shows the line of equality where $\mtwomin =
    M_2$; Systems near or above this threshold have candidate compact object
    companions as the more massive secondary is fainter than the primary.
    The lower dashed (blue) line shows the approximate hydrogen burning limit;
    Sources below this threshold are candidate substellar objects.
    The eight highlighted points (red) are shown below in \figurename
    s~\ref{fig:brown-dwarfs} and \ref{fig:compact-objects}.
    \textbf{Right:} The ratio of the primary stellar radius to the (projected)
    system orbital semi-major axis as a function of minimum mass ratio.
    Sources above the dashed (blue) line are likely interacting and may be
    photometrically variable.
    \label{fig:m2m1}
    }
\end{figure}

By construction, all sources in the \goldsample\ have primary stellar mass
estimates, $M_1$, from the \acronym{STARHORSE} catalog \citep{Queiroz:2019}.
For these systems, we can then also convert the inferred orbital parameters into
measurements of the minimum companion mass, \mtwomin\ (i.e., $M_2 \, \sin i$
where the unknown inclination $i$ is set to $90^\circ$).
\figurename~\ref{fig:m2m1} (left panel) shows these minimum companion mass
estimates as a function of the \acronym{STARHORSE} primary masses for all
sources in the \goldsample.
While the uncertainties in these quantities are not shown for most sources, the
eight highlighted systems (red markers with error bars) show typical values of
the errors on the masses (but note that the errors will be strongly correlated).
The two dashed (blue) lines show the approximate hydrogen burning limit (lower
horizontal line), and the upper curve shows the line of equality where the
minimum companion mass is equal to the primary mass.
Of these, 95 systems have $\mtwomin < 80~\mjup$: Some of these may be
high-inclination stellar-mass systems, but all should be considered brown dwarf
candidates.
Based on the quality cuts applied to define the parent sample (which should
remove sources with blended spectral lines), systems with $\mtwomin > M_1$
should not exist in the sample if the companion is luminous; The 40 systems with
$\mtwomin > M_1$ are therefore excellent candidate compact object companions and
will be discussed in a separate paper (Price-Whelan et al., in prep.).

% Notebook: Figure-gold-sample.ipynb

The right panel of \figurename~\ref{fig:m2m1} shows the ratio of the primary
stellar radius over the (projected) system semi-major axis as a function of
(minimum) mass ratio.
Here, the curved, dashed line shows an estimate of the Roche radius
\citep{Eggleton:1983}: Systems above this line are likely interacting.
One such system (\texttt{2M08160493+2858542}), indicated by the square (orange)
marker in this panel, appears to be strongly photometrically variable in data
from the \acronym{ASAS-SN} survey \citep{Shappee:2014, Jayasinghe:2019}, however
most other candidate interacting systems do not have \acronym{ASAS-SN} light
curves and could instead be followed up with \acronym{TESS} \citep{Ricker:2014}.

% Notebook: Figure-highmass-lowmass.ipynb
\begin{figure}[!t]
    \begin{center}
    \includegraphics[width=0.8\textwidth]{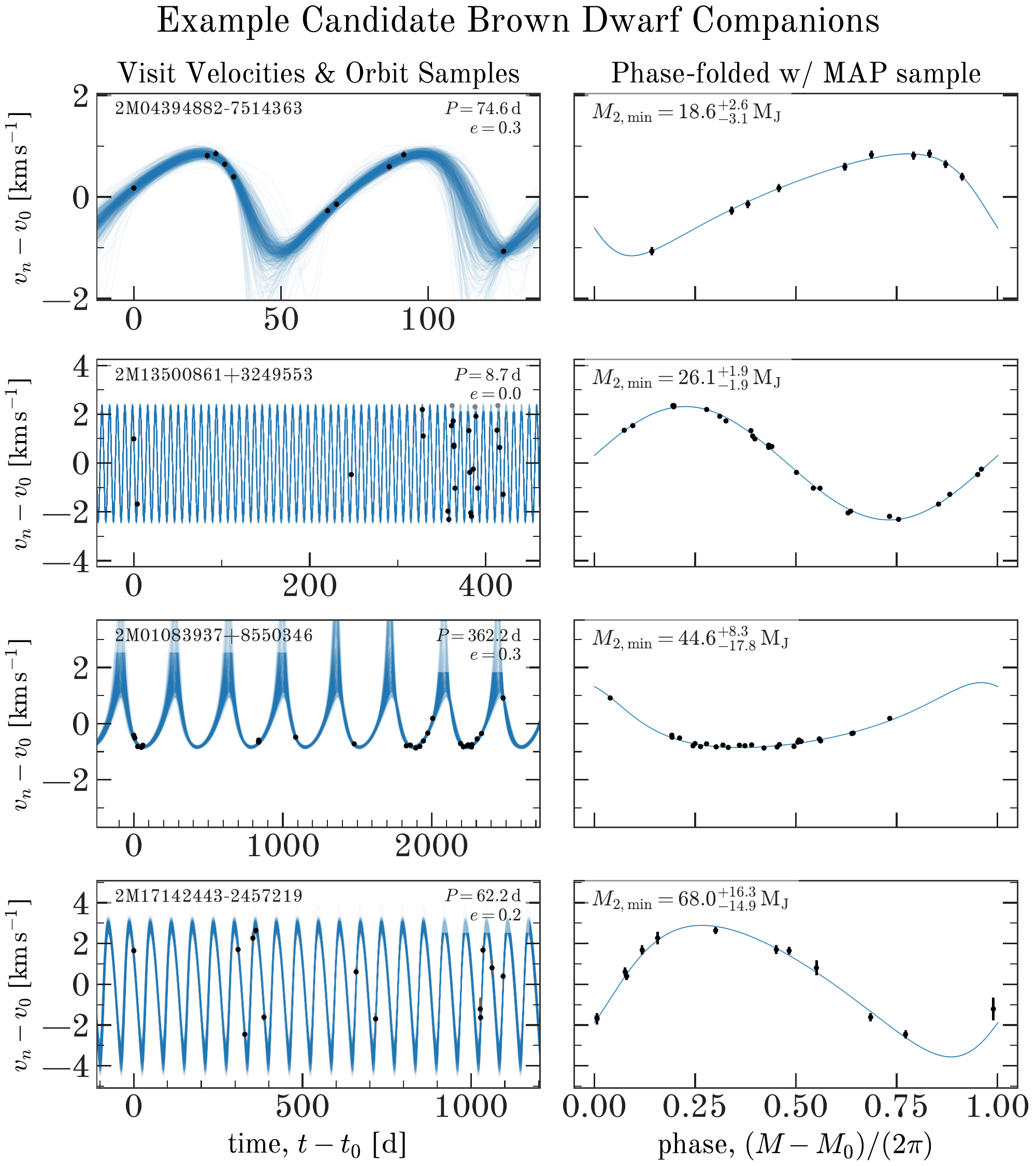}
    \end{center}
    \caption{%
    Example binary star systems from the \goldsample\ with low-mass companions
    that are candidate substellar objects.
    Each row is a different source (indicated in the left panels).
    The left panels show the raw visit velocity data (black markers)
    under-plotted with orbits computed from the posterior samples for each
    source (blue lines).
    The right panels show the same data phase-folded with the MAP orbit sample
    period and under-plotted with an orbit computed from the MAP sample.
    The minimum companion mass for each system is indicated in the right panels.
    \label{fig:brown-dwarfs}
    }
\end{figure}

\figurename~\ref{fig:brown-dwarfs} shows the radial velocity data (black
markers)---under-plotted (blue lines) with orbits computed from posterior
samples---for the four highlighted systems below the $80~\mjup$ line in
\figurename~\ref{fig:m2m1} (left).
The left panels show the time series, and the right panels show the same data
and orbits, but now phase-folded using the MAP period value.
The inferred minimum companion masses are indicated on each right panel.
These systems were chosen from a vetted subsample of all substellar companion
candidates to highlight systems with a range of companion masses,
eccentricities, and numbers of observations.

\begin{figure}[!t]
    \begin{center}
    \includegraphics[width=0.8\textwidth]{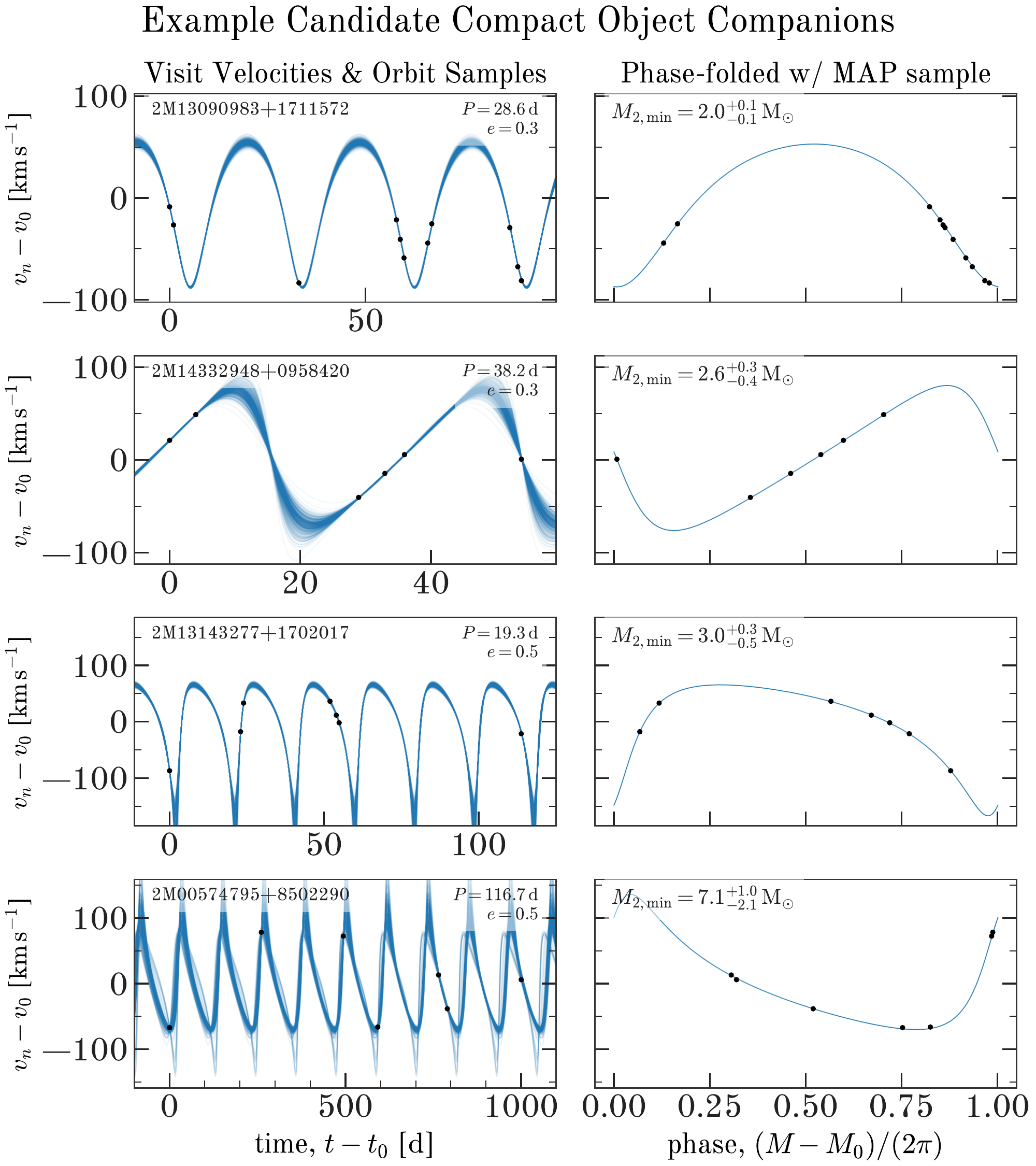}
    \end{center}
    \caption{%
    The same as \figurename~\ref{fig:brown-dwarfs}, but for example binary star
    systems from the \goldsample\ with faint, high-mass companions that are
    candidate compact objects.
    \label{fig:compact-objects}
    }
\end{figure}

\figurename~\ref{fig:compact-objects} shows the same, but for the four
highlighted systems above the $\mtwomin = M_1$ line in
\figurename~\ref{fig:m2m1}.
The companions in the systems shown in the top two rows are just barely
consistent with being high-mass neutron stars \citep[e.g.,][]{Cromartie:2019},
but the systems shown in the bottom two rows contain candidate non-interacting
black hole companions (these are discussed in more detail in the companion
paper; Price-Whelan et al., in prep.).
The recently discovered non-interacting black hole--giant star system (that made
use of \apogee\ data; \citealt{Thompson:2019}) does not appear in our candidates
because it only has 3 visit spectra with \apogee\ data alone and thus does not
pass the strict cuts we used to construct the \goldsample.

% Notebook: Figure-highmass-lowmass.ipynb

\subsection{Hierarchical inference of the eccentricity distribution}
\label{sec:hierarch-ecc}

Most of the results highlighted above make use of the catalogs created from
defining selections on the posterior samplings generated with \thejoker.
However, the real power in the individual system posterior samplings is that
they enable further \emph{hierarchical} modeling of binary-star population
properties without having to make hard cuts on the samples.
In future work, we will use these samplings to construct a joint model for the
period, eccentricity, and mass ratio distributions of binary stars as a function
of stellar parameters, but this will also require modeling the sample selection
function and our detection efficiency.
While the full hierarchical inference is out of scope for this \documentname,
here we demonstrate how this could be done using a simpler, toy problem:
Inferring the eccentricity distribution of long-period binary stars using a
parametric model.

We first select all $53\,790$ sources with samplings from \thejoker\ (or
subsequent MCMC) with $4000~\textrm{K} < \Teff < 7000~\textrm{K}$ and $-2 < \mh
< 0.5$ in order to select main sequence, FGK-type stars.
We then only keep $8\,599$ sources that have $>128$ samples with $P > 100~\dayd$
and $K > 1~\kms$, of which 158 sources have unimodal samplings (i.e., generated
with MCMC).
The second criteria (on velocity semi-ampitude, $K$) would not be necessary if
we instead constructed a mixture model with components to represent stars with
companions and the background population, but here, for simplicity, we instead
just make a cut on the posterior samplings.
For each $j$ source in this sample of binaries with FGK-type primary stars, we
then have $M_j$ (up to 512) samples in orbital eccentricity, $e_{jm}$.
We parametrize the eccentricity distribution using a beta distribution,
$B(a,b)$, and use the importance sampling trick to reweight the $e_{jm}$ samples
to infer the hyperparameters $(a, b)$ \citep{Hogg:2010}.

In detail, we would like to evaluate or generate samples from the posterior
probability distribution over the hyperparameters of the eccentricity
distribution
\begin{equation}
    p(a, b \given D) \propto p(D \given a, b) \, p(a, b)\label{eq:hierarch-post}
\end{equation}
where $D$ represents all visit data for all sources, i.e.,
\begin{equation}
    p(D \given a, b) = \prod_j p(D_j \given a, b)
\end{equation}
and $(a, b)$ are the parameters of our assumed beta distribution model.
There is no obvious way to evaluate this posterior probability---especially the
likelihood term---given the hyperparameters.
However, recall that we have samplings from the per-source posterior
distributions over eccentricity,
\begin{align}
    p(e_j \given D_j, \alpha_0) &\propto p(D_j \given e_j) \,
        p(e_j \given \alpha_0)\\
    e_{jm} &\sim p(e \given D_j, \alpha_0)
\end{align}
where $\alpha_0$ is meant to represent the parameters of our assumed ``interim''
prior over eccentricity that we used to generate the per-source samplings (i.e.,
see \tablename~\ref{tbl:prior}), and here ``$\sim$'' means ``is sampled from.''
Through math that is explained in more detail in other work
\citep[e.g.,][]{Hogg:2010, Price-Whelan:2018}, it turns out that the
hierarchical likelihood can be written as a sum over the ratio of probabilities
\begin{equation}
    p(D_j \given a, b) \approx \frac{\mathcal{Z}}{M_j} \,
        \sum_k^M \frac{p(e_{jm} \given a, b)}{p(e_{jm} \given \alpha_0)}
\end{equation}
where $\mathcal{Z}$ is a normalization constant.
In practice, we evaluate the hierarchical log-likelihood, $\ln p(D_j \given a,
b)$, as
\begin{equation}
    \ln p(D \given a, b) =
        \sum_j \left[\underset{m}{\textrm{logsumexp}}\left(
        \ln p(e_{jm} \given a, b) - \ln p(e_{jm} \given \alpha_0)\right)
        - \ln M_j\right]
        \quad .
\end{equation}

% Notebook: Figure-Long-P-eccentricity.ipynb
\begin{figure}[!t]
    \begin{center}
    \includegraphics[width=0.85\textwidth]{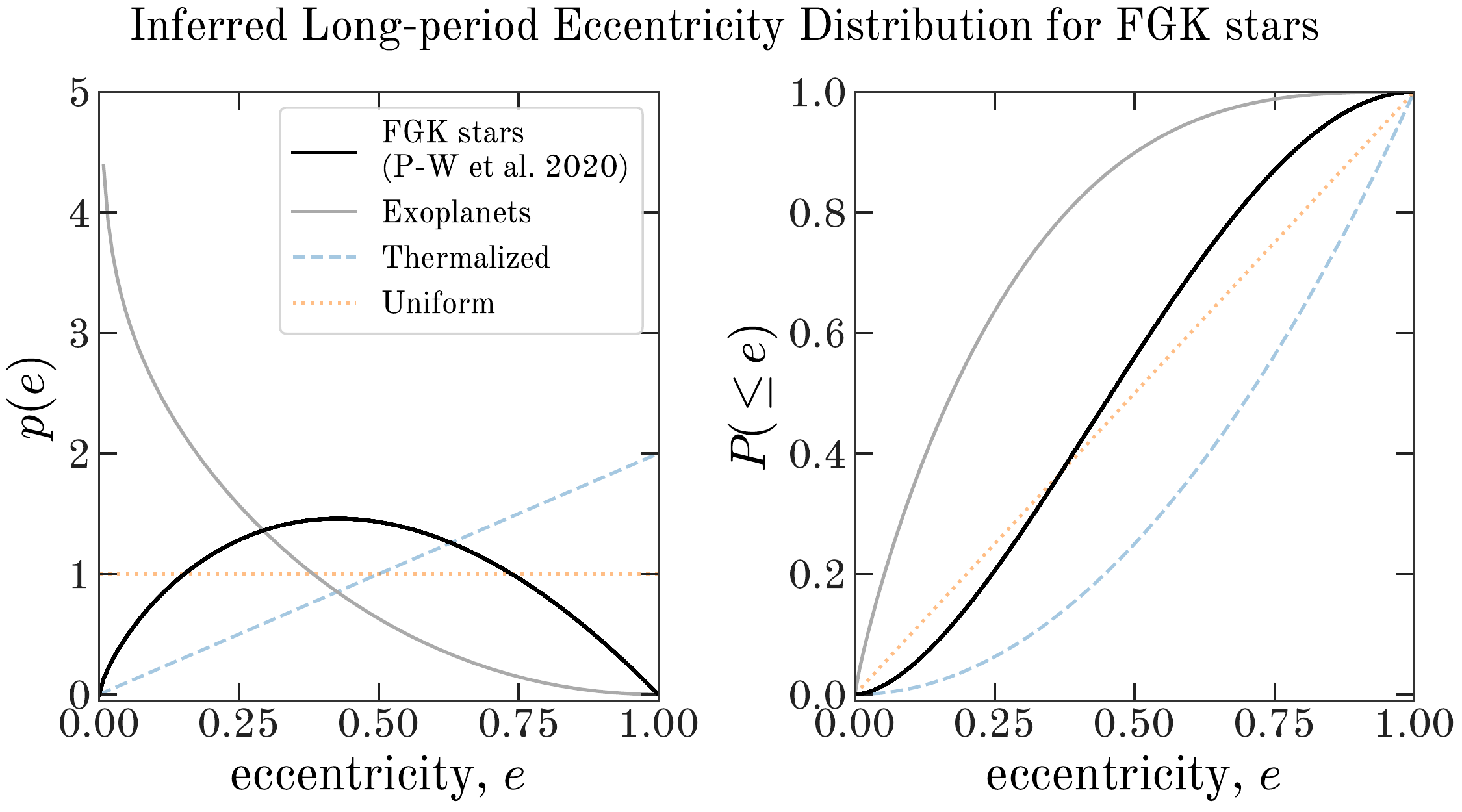}
    \end{center}
    \caption{%
    \textbf{Left:} The inferred eccentricity distribution for binary star
    systems with FGK-type primary stars (black line).
    We use the posterior sampings for $8\,599$ sources within a toy hierarchical
    model of the eccentricity distribution, represented as a beta distribution.
    We also plot common eccentricity distributions from the literature, such as
    the global exoplanet model from \cite[][; solid, gray line]{Kipping:2013},
    the (theoretical) thermalized population \citep[][; dashed, blue
    line]{Jeans:1919}, and a uniform distribution (dotted, orange line).
    \textbf{Right:} The same as the left panel, but showing cumulative
    distribution functions instead.
    \label{fig:eccdist}
    }
\end{figure}

With a method for evaluating the hierarchical likelihood
(\equationname~\ref{eq:hierarch-post}), we now just need to specify prior
probability distributions over the hyperparameters of the beta distribution $(a,
b)$.
For each parameter, we use a uniform distribution over the domain $(0.1, 10)$.
We implement this model, including the sums over the eccentricity samples,
within the context of a \texttt{pymc3} model and use the built-in NUTS sampler
to generate posterior samples in the parameters of the eccentricity
distribution.
We run the sampler for 1000 steps to tune, then run four chains in parallel,
each for an additional 2000 steps.
We assess convergence again by computing the Gelman-Rubin statistic, $\hat{R}$
and find that all parameters have converged samplings at the end of our run.

From these posterior samples, we find $a = 1.749 \pm 0.001$ and $b = 2.008 \pm
0.001$.
\figurename~\ref{fig:eccdist} shows the inferred eccentricity distribution
(black curve, left panel) and the corresponding cumulative distribution function
(right panel) along with some other eccentricity distributions from the
literature.
For example, the (global) exoplanet eccentricity distribution from
\citet{Kipping:2013}, the (theoretical) eccentricity distribution for a
thermalized population of binaries \citep{Jeans:1919}, and a uniform
distribution.
At long periods, binary star systems seem to have moderate eccentricities that
disfavor circular or very eccentric values.
This is in agreement with past studies that have focused on nearby samples of
solar-type stars with smaller sample sizes \citep[e.g.,][]{Duquennoy:1991,
Raghavan:2010}.
Note, however, that this should still be viewed as a demonstration: Future
analysis should assess the impact of selection effects and detection efficiency
on the inferred eccentricity distribution using these data.
We also emphasize that the extremely precise constraints we get on the
parameters of the beta distribution imply that we have enough data to complexify
the model, either by using non-parametric forms for the eccentricity
distribution to move away from rigid models or by parametrizing variations in
the distribution with stellar parameters.

\section{Discussion}\label{sec:discussion}

In what follows, we discuss some important caveats and considerations for
interpreting the results and using the catalogs described in this \documentname.

\subsection{Assumptions, caveats, and known failures}
\label{sec:caveats}

The assumptions that underlie the sampling procedure described above are
enumerated in \sectionname~3.1 in \cite{Price-Whelan:2018}.
We therefore briefly rephrase the most important implications of these
assumptions as a set of caveats and points of caution for any users of the
posterior samplings released with this \documentname.

First, despite allowing for a per-source, additive (in variance) extra
uncertainty parameter ($s$ in \tablename~\ref{tbl:prior}), we are very sensitive
to outliers, or, more generally, any visit velocities with strongly non-Gaussian
noise properties.
In \apogee, this might occur for blended sources, multiple-star systems with
more than one star that contribute significantly to infrared flux, or sources
with stellar parameters very beyond the grid of template spectra
\citep{Nidever:2015}.

Second, related to the first point, we make the strong assumption that all
sources are single-lined, i.e., that any binary companions do not contribute
significantly to the observed spectra.
This is wrong in general: Some systems will be double lined (sometimes only
subtly, but detectable, e.g., \citealt{El-Badry:2018}), and we will be biased in
cases where these sources are not properly filtered by the quality cuts done
above.
This also means that we will definitely miss systems with similar masses (and
especially ``twin'' binaries; \citealt{El-Badry:2019}).
We expect this assumption to be worse on the main sequence than on the giant
branch, because any lower-mass companion to a giant branch star will have a
luminosity hundreds or thousand times fainter.

Third, the adopted prior \pdf\ over systemic velocities, $v_0$, is reasonable
for considering all \apogee\ sources (i.e., for a mixture of kinematically
disk-like and halo-like sources), but may lead to biased posterior samplings for
sources in globular clusters or dwarf galaxies.
For such systems, it would be safer to re-run \thejoker\ with specialized priors
over systemic velocity for each individual host system.

Fourth, in this work, we assume that all systems are binary systems, so triples
or higher-order stellar multiplets will generally have incorrect samplings.
We will generate samplings for systems that are consistent with being
hierarchical triple-star systems, but this is beyond the scope of the
general-use catalog considered here.

Finally, related to the fourth point, we assume that all velocity variations
represent orbital motion (i.e., not pulsation or similar intrinsic stellar
phenomena).
As highlighted in \figurename~\ref{fig:binary-CMD}, this assumption is clearly
violated when $\logg \lesssim 1~\kms$ where stellar surface jitter masquerades
as orbital motion and leads to (a small amount of) contamination.

\subsection{Binary fraction trends with stellar properties}
\label{sec:binary-trends}

We find a number of interesting trends in the binary fraction as a function of
stellar parameters and chemical composition that have far-reaching implications.
For one, the binary fraction increases rapidly with decreasing metallicity (as
also noted recently by \citealt{Moe:2019, El-Badry:2019a}).
Beyond the implications discussed in \cite{Moe:2019}, this also motivates
appropriate investigation into how a large binary fraction could influence
inferred properties of dwarf galaxies and globular clusters.
For example, stellar binarity impacts velocity dispersion measurements, which
then contaminate estimates of dark matter masses in these systems
\citep[e.g.,][]{Aaronson:1987, Kouwenhoven:2008, Spencer:2017, Spencer:2018},
but also impact the long-term stability and dynamical evolution of compact
stellar systems \citep[e.g.,][]{Hut:1992, Sigurdsson:1993}.
A large binary fraction also impacts the chemical evolution of these systems
\citep[e.g.,][]{Eldridge:2008, Eldridge:2009}.

% While we detect many main sequence systems with short-period (i.e.,
% $P<50~\dayd$) companions, we do not observe systems with evolved-star members
% that have short-period companions that could have survived the common-envelope
% phase.
% Combined with the observation that the binary fraction rises steeply with
% stellar mass, this implies that an appreciable fraction of evolving stars (i.e.,
% the more massive stars on their way to the RGB or horizontal branch) likely
% merge with their companions.
% Do these mergers
% This , ... do the merger stars move to other places in the CMD? Bias stellar mass inferences for old stellar populations using SSP models?

% These trends undoubtedly also constrain binary star formation physics ...

We caution, however, that any real understanding of the binary fraction (in an
absolute sense) from the \emph{catalogs} produced here will require a model of
the \apogee\ selection function and for our detection efficiency.
Still, we clearly now have samples of binary stars large enough to begin placing
strong constraints on theories of binary star formation and evolution and their
impact on galactic stellar populations.

\subsection{Selection, completeness, false positives, and contamination}
\label{sec:completeness}

As cautioned above, to properly use our catalog of systems for performing
statistical tests (for example, to interpret the binary fraction values in
\figurename~\ref{fig:binary-CMD} in an absolute sense), it is critical to have
estimates of our detection efficiency or completeness as a function of binary
orbital parameters and stellar parameters.
While we have not provided estimates of the completeness, we do release the full
posterior samplings in orbital parameters for \emph{all} sources, along with
open-source software that could be used to construct these estimates.
In general, our completeness will likely be a strong function of velocity
semi-amplitude, period, and eccentricity (which implicitly imply functions of
companion mass, inclination, and separation).
It will also depend on mass ratio, as many sources with bright companions will
be dropped by the quality cuts imposed on the \apogee\ data.
As a consequence of this, the systems considered in this work are primarily
close binary systems with intermediate mass ratios.

One way to construct completeness estimates would be to repeat the analysis done
in this \documentname\ using simulated systems with known parameters.
The sampler used to generate the posterior orbit samplings is open source and
released as a \texttt{Python} package \citep{thejoker, Price-Whelan:2019a}.
The pipeline software used to define and analyze the \apogee\ parent sample is
open source and is also available as a \texttt{Python} package, \texttt{hq}
\citep{Price-Whelan:2019}.

% As for contamination, that is a hard problem, because it requires a
% model of the kinds of contaminants that might be present, which are
% (almost by definition) the things we don't know about.
% However, one good example is the asteroseismic mode contamination
% that appears to be enetering our sample at the top of the RGB.
% This is a kind of contamination that could certainly be modeled, with a
% mechanical model of asteroseismic surface motions and our open-source code.
% That's also out of scope, but an interesting, valuable, and straightforward
% project.

\subsection{Below-threshold searches}
\label{sec:subthreshold}

Another problem with a traditional catalog of detected binary companions (like
ours) is that there are many sources in which the hypothesis of a single star is
clearly rejected, but not strong enough to qualify as a reliable binary
``detection.''
A user who has informative data for a source that would take a sub-threshold
source above threshold (in terms of our criteria for including a source in the
catalog) has no recourse if only given the rigidly-thresholded catalog entries.
Even for well detected binary systems, in a traditional catalog of orbital
parameters, there is no simple way to combine the results with new data to
improve or adjust parameter estimates.
The samplings provided here can be used to solve these problems; They can be
used to combine the \apogee\ information with new data, without requiring an
\textsl{ab initio} re-analysis of the original \apogee\ data.
This can be done by repeating the rejection sampling step used in \thejoker\ to
generate posterior samples, but replacing the prior samples with the posterior
samples generated with the \apogee\ data alone and replacing the full marginal
likelihood evaluation with the marginal likelihood of the new data computed at
each sample.

In detail, following the notation in Appendix~\ref{app:marginal-likelihood},
converting the posterior samples released here into posterior samples over the
\apogee\ data, $D_1$, \textsl{plus} some new velocity data, $D_2$, requires
generating samples from the posterior \pdf
\begin{align}
    p(\bs{\theta} \given D_1, D_2) &\propto
        p(D_1 \given \bs{\theta}) \, p(D_2 \given \bs{\theta}) \,
        p(\bs{\theta})\label{eq:postupdate1}\\
        &\propto p(D_2 \given \bs{\theta}) \,
        \left[ p(D_1 \given \bs{\theta}) \, p(\bs{\theta}) \right]
        \label{eq:postupdate2}
\end{align}
where we have assumed that the new data are independent of the \apogee\ data.
Note that the terms in brackets in \equationname~\ref{eq:postupdate2} are
proportional to the posterior probability of the parameters given only the
\apogee\ data, i.e., the distribution we have generated samples from using
\thejoker.
We can therefore use the posterior samples generated from the \thejoker\ with
the \apogee\ data to rejection-sample using the new marginal likelihood, $p(D_2
\given \bs{\theta})$, to generate samples from $p(\bs{\theta} \given D_1, D_2)$.

The ability to perform sub-threshold searches or add external information is
limited by detailed shape of the posterior \pdf\ and the number of samples we
deliver (here, \Kminval).
If the new data are highly informative, or favor a mode in the posterior \pdf\
that is not well sampled, the posterior samplings we deliver will not be dense
enough to provide support for the updated posterior \pdf.
This puts limitations on the scope of sub-threshold searches, but at least they
are possible with these outputs, in some regime of applicability.

\section{Conclusions}

Our key results and conclusions are summarized below:
\begin{description}
    \item[The close binary fraction depends on the stellar parameters of the
    primary star] \figurename~\ref{fig:binary-CMD} shows a simple estimate of
    the close binary fraction over the color--magnitude diagram, utilizing all
    \nsources\ \apogee\ sources used in this work. There is a notable dearth of
    close companions around the red clump, where companions may have been
    engulfed when the primary star ascended the upper giant branch. We also
    rediscover trends in the binary fraction with effective temperature along
    the main sequence (i.e., stellar mass), and show that the binary fraction on
    the giant branch depends on the surface gravity (i.e., surface size) of the
    primary, also indicating that companion engulfment is an important outcome
    of close binary star evolution.
    \item[The binary fraction is anti-correlated with metallicity] We find that
    the binary fraction decreases linearly with bulk metallicity with a slope of
    $-0.1~\textrm{dex}^{-1}$ over the domain $-1 \lesssim \mh \lesssim 0.4$, as
    shown in \figurename~\ref{fig:binfrac-mh}.
    \item[We detect the clear signature of tidal circularization in field main
    sequence and red giant branch stars] \figurename~\ref{fig:Plogg} (left)
    shows inferred periods and eccentricities for systems with
    uniquely-determined orbits in the \goldsample.
    The abundances of low-eccentricity sources at short periods is a
    manifestation of tidal circularization, which should operate at longer
    periods for larger, more convective (i.e., giant branch) stars.
    \item[We identify 95 candidate brown dwarf companions] Using simplistic cuts
    on the \goldsample\ of high-quality sources with unimodal posterior
    samplings, we identify candidate brown dwarf companions by selecting sources
    with median minimum companion mass (\mtwomin) values below the hydrogen
    burning limit, $\mtwomin < 80~\mjup$. \figurename~\ref{fig:m2m1} (left)
    shows these sources as points below the horizontal dashed line. We highlight
    a few of these systems in \figurename~\ref{fig:brown-dwarfs}.
    \item[We identify 40 candidate non-interacting compact object companions]
    Again using simplistic cuts on the \goldsample\ of high-quality sources with
    unimodal posterior samplings, we identify candidate compact object
    companions by selecting sources with $\mtwomin > M_1$.
    \figurename~\ref{fig:m2m1} (left) shows these sources as points above the
    upper, curved dashed line. We highlight a few of these systems in
    \figurename~\ref{fig:compact-objects}.
    \item[The binary-star eccentricity distribution is peaked at
    moderate eccentricities] We execute a toy hierarchical inference using the
    posterior samplings for $\sim 8\,600$ FGK-type main sequence stars to infer
    the eccentricity distribution of long-period ($P>100~\dayd$),
    intermediate-mass main-sequence star binary systems. By representing the
    distribution using a beta distribution, we derive precise posterior
    constraints on the parameters using this hierarchical model and find $a =
    1.749 \pm 0.001$ and $b = 2.008 \pm 0.001$. \figurename~\ref{fig:eccdist}
    shows our inferred eccentricity distribution, indicating that long-period
    binary star systems prefer moderate eccentricities.
    \item[We release a sample of $20\,000$ binary star systems and posterior
    samplings over orbital parameters for $\nsources$ \apogee\ sources] Finally,
    we release a catalog of \nbinary\ high-confidence binary star systems
    (\tablename~\ref{tbl:metadata}). The majority of these systems have poorly
    constrained orbital parameters, but we release posterior samplings over
    these parameters for all \nsources\ \apogee\ sources to enable other
    probabilistic inferences with these data. We also define and release a
    \goldsample\ containing \ngold\ systems with high-quality, unimodal
    posterior samplings that can be used and summarized more simply.
\end{description}

\acknowledgements

It is a pleasure to thank
Borja Anguiano (Virginia),
Carles Badenes (Pittsburgh),
Evan Bauer (KITP),
Lars Bildsten (KITP),
Dan Foreman-Mackey (Flatiron),
Will Farr (Flatiron),
Marla Geha (Yale),
and Adam Jermyn (Flatiron) for useful discussions.
APW acknowledgements support and space from the Max-Planck-Institut f\"ur
Astronomie during initial work on this project.
PMF acknowledge support for this research from the National Science Foundation
(AST-1311835 \& AST-1715662).
JGF-T is supported by FONDECYT No. 3180210 and Becas Iberoam\'erica Investigador
2019, Banco Santander Chile.
DAGH acknowledges support from the State Research Agency (AEI) of the Spanish
Ministry of Science, Innovation and Universities (MCIU) and the European
Regional Development Fund (FEDER) under grant AYA2017-88254-P.
SH is supported by an NSF Astronomy and Astrophysics Postdoctoral Fellowship
under award AST-1801940.
% We thank the anonymous referee for constructive comments that improved this
% manuscript.

Funding for the Sloan Digital Sky Survey IV has been provided by the Alfred P.
Sloan Foundation, the U.S. Department of Energy Office of Science, and the
Participating Institutions. SDSS-IV acknowledges support and resources from the
Center for High-Performance Computing at the University of Utah. The SDSS web
site is www.sdss.org.

SDSS-IV is managed by the Astrophysical Research Consortium for the
Participating Institutions of the SDSS Collaboration including the Brazilian
Participation Group, the Carnegie Institution for Science, Carnegie Mellon
University, the Chilean Participation Group, the French Participation Group,
Harvard-Smithsonian Center for Astrophysics, Instituto de Astrof\'isica de
Canarias, The Johns Hopkins University, Kavli Institute for the Physics and
Mathematics of the Universe (IPMU) / University of Tokyo, Lawrence Berkeley
National Laboratory, Leibniz Institut f\"ur Astrophysik Potsdam (AIP),
Max-Planck-Institut f\"ur Astronomie (MPIA Heidelberg), Max-Planck-Institut
f\"ur Astrophysik (MPA Garching), Max-Planck-Institut f\"ur Extraterrestrische
Physik (MPE), National Astronomical Observatories of China, New Mexico State
University, New York University, University of Notre Dame, Observat\'ario
Nacional / MCTI, The Ohio State University, Pennsylvania State University,
Shanghai Astronomical Observatory, United Kingdom Participation Group,
Universidad Nacional Aut\'onoma de M\'exico, University of Arizona, University
of Colorado Boulder, University of Oxford, University of Portsmouth, University
of Utah, University of Virginia, University of Washington, University of
Wisconsin, Vanderbilt University, and Yale University.

\software{
    Astropy \citep{astropy, astropy:2018},
    apred \citep{Nidever:2015},
    ASPCAP \citep{ASPCAP},
    exoplanet \citep{exoplanet:exoplanet},
    gala \citep{gala},
    IPython \citep{ipython},
    numpy \citep{numpy},
    pymc3 \citep{Salvatier2016},
    schwimmbad \citep{schwimmbad:2017},
    scipy \citep{scipy},
    theano \citep{theano},
    thejoker \citep{thejoker, Price-Whelan:2019a}
}

\appendix

\section{Update to the marginal likelihood expression for \thejoker}
\label{app:marginal-likelihood}

As noted above (see \sectionname~\ref{sec:joker-update}), the assumptions in
\cite{thejoker} that lead to the simplified form of the marginal likelihood
(\equationname~11 in \citealt{thejoker}) are not valid.
We therefore here re-derive the marginal likelihood that forms the basis of the
implementation of \thejoker\ used in this work.

For each \apogee\ source, we have a set of $N$ radial velocity measurements
(visits) $v_n$ at times $t_n$ with uncertainties $\sigma_n$.
Under the assumption that the source is in a binary star system, our model for
the true radial velocity of the source at any time is given by
\equationname~\ref{eq:kepler} above.
Our goal (as in \citealt{thejoker}) is to analytically marginalize over the
linear parameters in \equationname~\ref{eq:kepler}---$(K, v_0)$---under the
assumption that the uncertainties on each radial velocity measurement are
Gaussian and independent.
To do this, we must write down expressions for the likelihood, and for the prior
probability distribution over the parameters that we will be marginalizing over
(i.e., the linear parameters).
We have already made the assumption that our likelihood has a Gaussian form, so
to do this marginalization conveniently and analytically, we additionally assume
that the prior \pdf\ also has a Gaussian form.
Under these assumptions, the solution to this marginalization is given in
Hogg et al. (in prep.).

To see the relation between the specific problem solved by \thejoker\ and the
derivation in Hogg et al. (in prep.), it will be convenient to repackage our
data and linear parameters as
\begin{align}
    \vec{y} &= \transpose{(v_1, v_2, \ldots, v_N)}\\
    \vec{x} &= \transpose{(K, v_0)}\\
    \mat{C} &=
        \begin{pmatrix}
            \sigma_1^2 & 0 & \cdots & 0\\
            0 & \sigma_2^2 & \cdots & 0\\
             &  & \ddots & \\
            0 & 0 & \cdots & \sigma_N^2\\
        \end{pmatrix}
\end{align}
and to assume that the mean and covariance matrix of the prior over the linear
parameters are given by $\vec{\mu}, \mat{\Lambda}$, respectively.
Given a set of nonlinear parameters $\vec{\theta} = (P, e, \omega, M_0)$, we
will also need to define a design matrix, $\mat{M}$,
\begin{equation}
    \mat{M} =
        \begin{pmatrix}
            \zeta(t_1 \,;\, \vec{\theta}) & 1\\
            \zeta(t_2 \,;\, \vec{\theta}) & 1\\
            \vdots & \vdots\\
            \zeta(t_N \,;\, \vec{\theta}) & 1
        \end{pmatrix}
\end{equation}
where $\zeta(\cdot)$ is given by \equationname~\ref{eq:zeta}.
With these assumptions, the marginalized likelihood, $Q$, for a source, given
nonlinear parameters $\vec{\theta}$, can be written
\begin{align}
    Q &= \int \dd \vec{x} \,
        \norm(\vec{y} \given \mat{M}\cdot \vec{x}, \mat{C}) \,
        \norm(\vec{x} \given \vec{\mu}, \mat{\Lambda}) \\
    &= \int \dd \vec{x} \,
        \norm(\vec{x} \given \vec{a}, \mat{A}) \,
        \norm(\vec{y} \given \vec{b}, \mat{B})\\
    &= \norm(\vec{y} \given \vec{b}, \mat{B})\\
    \vec{b} &= \mat{M} \cdot \vec{\mu}\\
    \mat{B} &= \mat{C} + \mat{M} \cdot \mat{\Lambda} \cdot \transpose{\mat{M}}
\end{align}
where the integral becomes simple because the second normal distribution,
$\norm(\vec{y} \given \vec{b}, \mat{B})$, does not depend on the integration
variables, $\vec{x}$, and the integral over the first normal distribution is 1
Hogg et al. (in prep.).
A final point that is relevant to additional enhancements discussed in
\sectionname~\ref{sec:joker-update} (and is exploited in this \documentname) is
to note that $\vec{\mu}$ and $\mat{\Lambda}$ \emph{can} depend on the nonlinear
parameters.

\section{Data tables}
\label{sec:datatables}

The primary data product released with this \documentname\ are the posterior
samplings generated for each of \nsources\ sources in \apogee\ \dr{16}.
However, we also compute summary information and statistics about these
samplings and provide these data in \tablename~\ref{tbl:metadata}.
We also define a \goldsample\ of high-quality, uniquely-solved binary star
systems (see \sectionname~\ref{sec:gold-sample}) and release summary information
along with cross-matched data from \gaia\ \dr{2} and the \acronym{STARHORSE}
catalog of stellar parameters in \tablename~\ref{tbl:goldsample}.

\begin{table}[ht]
    \footnotesize
    \centering
    \begin{tabular}{l|l|p{6.6cm}}
        \multicolumn{3}{c}{\textbf{Metadata for all sources in the parent sample}} \\
        \hline
        Column name & Unit / format & Description \\
        \hline
        \texttt{APOGEE\_ID}                          &                        & \apogee\ source identifier\\
        \texttt{n\_visits}                           &                        & number of visits that pass our quality cuts\\
        \texttt{MAP\_P}                              & $\mathrm{d}$           & $P$, orbital period\\
        \texttt{MAP\_P\_err}                         & $\mathrm{d}$           & \\
        \texttt{MAP\_e}                              &                        & $e$, orbital eccentricity\\
        \texttt{MAP\_e\_err}                         &                        & \\
        \texttt{MAP\_omega}                          & $\mathrm{rad}$         & $\omega$, argument of pericenter\\
        \texttt{MAP\_omega\_err}                     & $\mathrm{rad}$         & \\
        \texttt{MAP\_M0}                             & $\mathrm{rad}$         & $M_0$, phase at reference epoch\\
        \texttt{MAP\_M0\_err}                        & $\mathrm{rad}$         & \\
        \texttt{MAP\_K}                              & $\mathrm{km\,s^{-1}}$  & $K$, velocity semi-amplitude\\
        \texttt{MAP\_K\_err}                         & $\mathrm{km\,s^{-1}}$  & \\
        \texttt{MAP\_v0}                             & $\mathrm{km\,s^{-1}}$  & $v_0$, systemic velocity\\
        \texttt{MAP\_v0\_err}                        & $\mathrm{km\,s^{-1}}$  & \\
        \texttt{MAP\_s}                              & $\mathrm{km\,s^{-1}}$  & $s$, excess variance parameter\\
        \texttt{MAP\_s\_err}                         & $\mathrm{km\,s^{-1}}$  & \\
        \texttt{t0\_bmjd}                            &                        & reference epoch (Barycentric MJD)\\
        \texttt{baseline}                            & $\mathrm{d}$           & the time baseline of the visits\\
        \texttt{MAP\_ln\_likelihood}                 &                        & the log-marginal-likelihood value of the MAP sample\\
        \texttt{MAP\_ln\_prior}                      &                        & the log-prior value of the MAP sample\\
        \texttt{max\_unmarginalized\_ln\_likelihood} &                        & the maximum value of the unmarginalized likelihood\\
        \texttt{max\_phase\_gap}                     &                        & the maximum gap in phase of the data for the MAP sample\\
        \texttt{periods\_spanned}                    &                        & the number of (MAP) periods spanned by the data \\
        \texttt{phase\_coverage}                     &                        & the phase-coverage of the data folded on the MAP period\\
        \texttt{phase\_coverage\_per\_period}        &                        & the maximum number of data points within a single MAP period\\
        \texttt{unimodal}                            & \texttt{boolean}       & True if the sampling is unimodal in period\\
        \texttt{joker\_completed}                    & \texttt{boolean}       & True if the source has \Kminval\ samples from \thejoker\\
        \texttt{mcmc\_completed}                     & \texttt{boolean}       & True if the source has \Kminval\ samples from MCMC\\
        \texttt{mcmc\_success}                       & \texttt{boolean}       & True if the MCMC sampling converged\\
        \texttt{gelman\_rubin\_max}                  &                        & The maximum (over parameters) value of $\hat{R}$\\
        \texttt{robust\_constant\_ln\_likelihood}    &                        & The maximum log-likelihood value of a robust constant-velocity fit to the visit data\\
        \texttt{binary\_catalog}                     & \texttt{boolean}       & True if the source is in the binary catalog (\sectionname~\ref{sec:catalog})\\
        \hline
        \multicolumn{3}{c}{\textit{(\nsources\ rows)}}
    \end{tabular}
    \caption{Description of the metadata table containing summary information
    for all \apogee\ sources in the parent sample.
    All columns ending in \texttt{\_err} are estimates of the standard-deviation
    of the posterior samples around the MAP sample, but only computed for
    unimodal samplings.
    }
    \label{tbl:metadata}
\end{table}

\begin{table}[ht]
    \footnotesize
    \centering
    \begin{tabular}{l|l|p{6.5cm}}
        \multicolumn{3}{c}{\textbf{Metadata for the \goldsample}} \\
        \hline
        Column name & Unit / format & Description \\
        \hline
        \texttt{APOGEE\_ID} & & \apogee\ source identifier\\
        \texttt{mass}        & $\mathrm{M_{\odot}}$ & mass of the primary, from the \acronym{STARHORSE} catalog\\
        \texttt{mass\_err}   & $\mathrm{M_{\odot}}$ & uncertainty on the primary mass\\
        \texttt{m2\_min\_1}  & $\mathrm{M_{\odot}}$ & 1st percentile value of \mtwomin\\
        \texttt{m2\_min\_5}  & $\mathrm{M_{\odot}}$ & 5th percentile value of \mtwomin\\
        \texttt{m2\_min\_16} & $\mathrm{M_{\odot}}$ & 16th percentile value of \mtwomin\\
        \texttt{m2\_min\_50} & $\mathrm{M_{\odot}}$ & median value of \mtwomin\\
        \texttt{m2\_min\_84} & $\mathrm{M_{\odot}}$ & 84th percentile value of \mtwomin\\
        \texttt{m2\_min\_95} & $\mathrm{M_{\odot}}$ & 95th percentile value of \mtwomin\\
        \texttt{m2\_min\_99} & $\mathrm{M_{\odot}}$ & 99th percentile value of \mtwomin\\
        \vdots & & all columns from \tablename~\ref{tbl:metadata}\\
        \vdots & & all columns from \apogee\ \texttt{allStar} file\\
        \vdots & & all columns from \gaia\ DR2 \texttt{gaia\_source} table\\
        \hline
        \multicolumn{3}{c}{\textit{(\ngold\ rows)}}
    \end{tabular}
    \caption{Description of the data table containing summary information
    for all sources in the \goldsample.
    }
    \label{tbl:goldsample}
\end{table}

\bibliographystyle{aasjournal}
\bibliography{dr16vac}

\end{document}